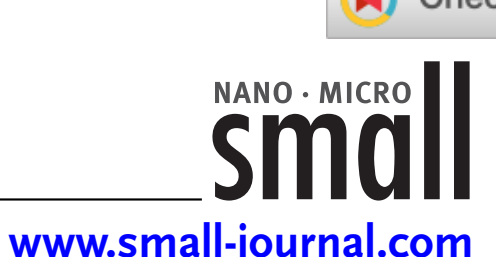

# Quantum Transport in Nitrogen-Doped Nanoporous Graphenes

*Gaetano Calogero, Isaac Alcón,* Alan E. Anaya Morales, Nick Papior, Pol Febrer, Aron W. Cummings, Miguel Pruneda, Stephan Roche, and Mads Brandbyge*

**Bottom-up on-surface synthesized nanoporous graphenes (NPGs), realized as 2D arrays of laterally covalently bonded $\pi$-conjugated graphene nanoribbons (GNRs), are a family of carbon nanomaterials that are receiving increasing attention for nanoelectronics and biosensing. Recently, a so-called hybrid-NPG (hNPG) is synthesized, featuring an alternating sequence of doped and non-doped GNRs, resulting in a band staggering effect in its electronic structure. Such a feature is appealing for photo-catalysis, photovoltaics and even carbon nanocircuitry. However, to date, little is known about the transport properties of hNPG and its derivatives, which is key for most applications. Here, via Green's functions simulations, the quantum transport properties of hNPGs are studied. It is found that injected carriers in hNPG spread laterally through a number of GNRs, though such spreading may take place exclusively through GNRs of one type (doped or non-doped). A simple model is proposed to discern the key parameters determining the electronic propagation in hNPGs and explore alternative hNPG designs to control the spreading/confinement and anisotropy of charge transport in these systems. For one such design, it is found that it is possible to send directed electric signals with sub-nanometer precision for as long as one micrometer – a result first reported for any NPG.**

## 1. Introduction

The power of bottom-up on-surface synthesis to realize virtually any carbon nanostructure has been demonstrated during the last two decades.[1–3] While a number of nanomaterials have been reported in this way, including 2D covalent organic frameworks[1] and molecular nanographenes,[4,5] only graphene nanoribbons (GNRs) have acquired major attention for nanoelectronics.[6,7] This is partly due to their fully $\pi$-conjugated nature and semiconducting band structure, which opens the door to control current flow through such nanomaterials via electrostatic means. In spite of the significant amount of research devoted to GNR-based devices, there are still major challenges, such as controlling the length of bottom-up synthesized GNRs or their arrangement within solid-state devices.[7] The so-called nanoporous graphenes (NPGs) are GNR-based 2D materials that could solve some of

G. Calogero
National Research Council
Institute for Microelectronics and Microsystems (CNR-IMM)
Zona Industriale, Strada VIII, 5, Catania 95121, Italy

I. Alcón, P. Febrer, A. W. Cummings, M. Pruneda, S. Roche
Catalan Institute of Nanoscience and Nanotechnology (ICN2)
CSIC and BIST, Campus UAB
Bellaterra, Barcelona 08193, Spain
E-mail: ialcon@ub.edu

I. Alcón
Institute of Theoretical and Computational Chemistry
Department of Materials Science and Physical Chemistry
Universitat de Barcelona
C/ de Martí i Franquès, 1–11, Les Corts, Barcelona 08028, Spain

A. E. A. Morales, M. Brandbyge
Department of Physics
Technical University of Denmark
Kongens Lyngby DK-2800, Denmark

N. Papior
Computing Center
Technical University of Denmark
Kongens Lyngby DK-2800, Denmark

M. Pruneda
Nanomaterials and Nanotechnology Research Centre (CINN-CSIC)
Universidad de Oviedo (UO)
Principado de Asturias, El Entrego 33940, Spain

S. Roche
ICREA
Institució Catalana de Recerca i Estudis Avançats
Barcelona 08070, Spain

The ORCID identification number(s) for the author(s) of this article can be found under https://doi.org/10.1002/smll.202508850

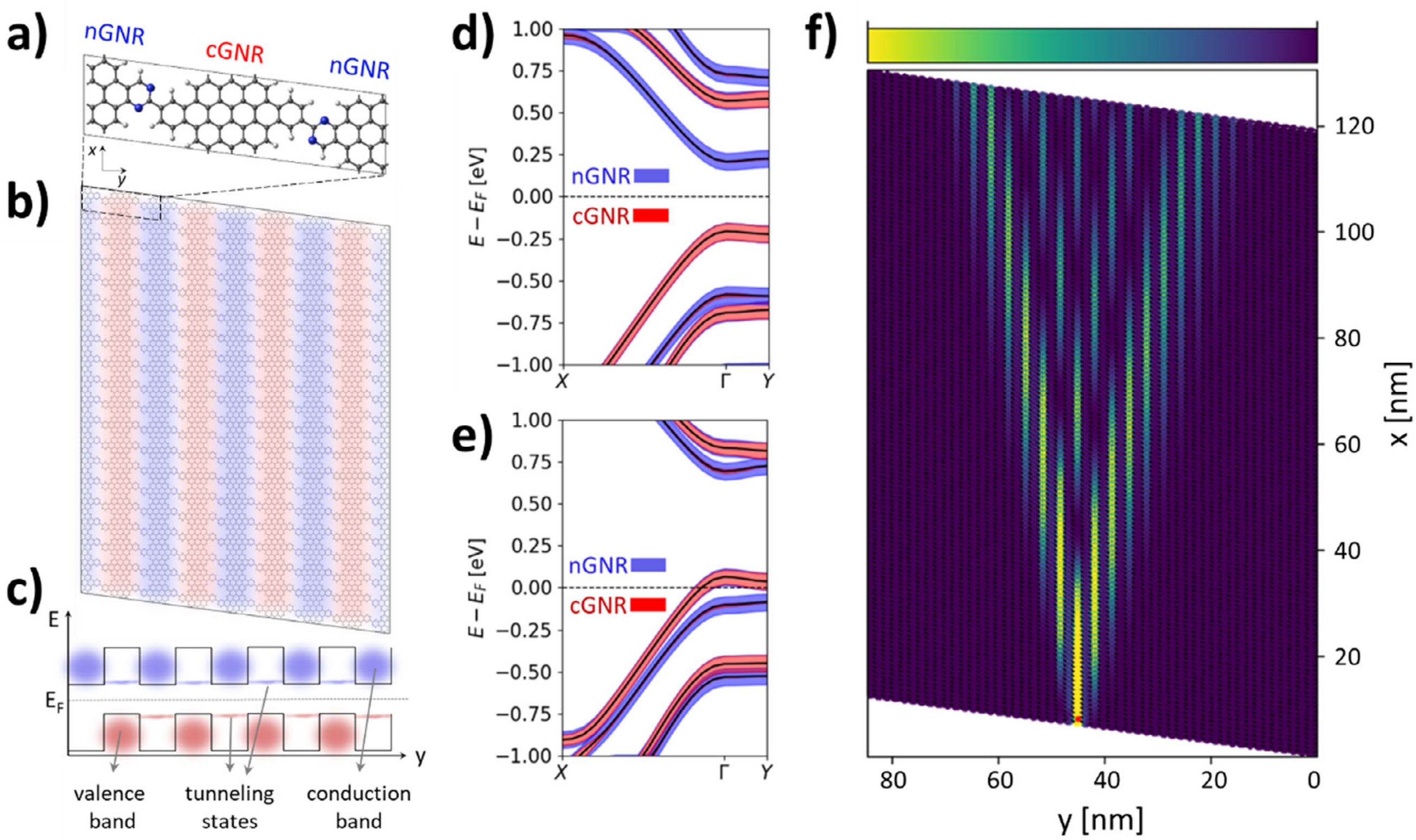

**Figure 1.** a) Optimized atomic structure of hNPG (C: gray, N: blue, H: white). b) Medium-scale atomistic model of hNPG. c) Outline of the type II band staggering in hNPG, where the valence and conduction bands are highlighted in red and blue, respectively. Tunneling states are also highlighted with the same color scheme as the GNR type. d) GNR-resolved fat band structure (cGNR: red; nGNR: blue) for the charge-neutral system and e) the $-0.89 \cdot 10^{13}$ e cm$^{-2}$ p-gated system ($-0.25$ e cell$^{-1}$). f) Bond transmission maps of an ≈85 × 122 nm$^2$ hNPG sample gated as in e), upon locally contacting the central cGNR at the bottom of the device (see small red dot). Areas with high bond transmission are shown in bright yellow, those with low or no bond transmission are in dark (color bar ranging from 0 to 0.045).

these issues. NPGs, first reported in 2018,[8] are 2D covalent arrays of laterally bonded GNRs, and so they may be regarded as a 2D net of electronically coupled 1D nanochannels for charge carriers.[9] Different types of NPGs have been synthesized since then, by varying the GNR type (including armchair, chevron-like, etc.)[10,11] or the covalent linkage between the ribbons, including direct C─C bonding or molecular bridges of varying length.[12,13] Additionally, the tremendous progress on fabricating GNR-based heterojunctions[14–18] should possibly be integrated in NPGs in the near future, further expanding the available set of NPG nanoarchitectures. While NPG research is still in the early stages, some researchers have managed to integrate it into solid-state devices.[8,19] This growing interest has recently been summarized in different review articles focusing on NPG synthesis[9,20] and quantum transport characteristics.[21]

One of the most recent members of the NPG family of materials is the so-called hybrid-NPG (hNPG for short).[22] The hNPG is formed by 7–13 armchair GNRs (7-13-AGNRs), as for the original NPG,[8] but every other GNR in hNPG is nitrogen-doped (N-doped), making the hNPG a 2D array of alternating N-doped GNRs (nGNRs hereafter) and non-doped GNRs (cGNRs), as depicted in **Figure** 1a,b. Such chemical alternation gives rise to a type II band-staggering effect,[22] where the valence band of the material originates from cGNRs and the conduction band originates from the nGNR array (see Figure 1c). As such, hNPG is a lateral heterojunction with atomically sharp band offsets alternating on the nanometer scale, which makes it a relevant platform for applications in optoelectronics, photovoltaics, photocatalysis, and selective ion sieving.[22] However, hNPG also displays in-gap tunnelling states in the energy range spanned by the valence (conduction) band that couple all cGNRs (nGNRs) through nGNRs (cGNRs), as schematically represented in Figure 1c (see "tunneling states"). The existence of these in-gap tunneling states could hinder the lateral confinement of electron propagation and limit spatial control.

In this work, we study the quantum transport properties of hNPG and its derivatives. We do this via density functional theory (DFT) combined with the Green's function (GF) formalism. Concretely, we simulate the application of gates to set the Fermi energy ($E_F$) either within the valence or conduction band of hNPG, which allows us to evaluate the behavior of injected currents upon locally probing either a cGNR or a nGNR. We find that the band staggering effect does not prevent the lateral spreading of locally injected currents, which we associate with the presence of tunneling states that become more pronounced upon gating. Inspired by former work on Talbot interference in NPGs,[23] we develop an analytical model to describe the current propagation patterns emerging in hNPG, in analogy to photons propagating in coupled waveguides. We then investigate two design strategies toward transport confinement in hNPG-based systems. First, inspired by prior studies,[22] we consider alternative hNPG lateral distributions where a cGNR is separated from the next cGNR by an increasing number of nGNRs. Second, we combine the alternating doping of hNPG with the use of phenyl rings to bridge neighboring GNRs in meta-configuration, which, as previously studied in depth,[13,24–26] leads to destructive quantum interference (QI) effects and strong lateral confinement of charge carriers. While both strategies significantly reduce the spreading of currents as compared to hNPG, we find that only the second route leads to full confinement of injected currents within the single contacted GNR. In the second scenario, we demonstrate that by tuning the gate voltage, the injected charge carriers can stay within a single GNR for more than one micrometer away from the injection point. Such a level of control over current flow at the nanoscale represents a clear improvement over QI-based strategies alone and demonstrates a synergistic effect upon combining QI and band-staggering engineering.

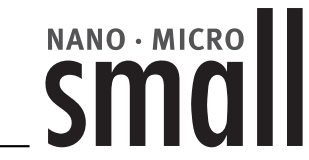

## 2. Results and Discussion

Figure 1a displays the optimized structure of hNPG, built as a lateral heterostructure of (N-doped) nGNRs and (non-doped) cGNRs laterally connected via single C─C bonds. As previously mentioned, such N-doping leads to an energetic stabilization of nGNR states with respect to cGNR states.[22] As a consequence, the valence band consists nearly entirely of cGNR orbitals, whereas the conduction band is made up of nGNR orbitals, as shown by the GNR-resolved fat band structure in Figure 1d. Similar to other NPGs, both the valence and conduction bands are primarily dispersive along the GNRs ($x$-direction; $\Gamma \rightarrow X$), indicative of their conductive character. However, in Figure 1d we also see a non-negligible band dispersion across the GNRs ($\Gamma \rightarrow Y$ segment), which indicates a certain degree of electronic coupling between GNRs of the same type. As previously reported,[22] such finite coupling originates from the presence of tunnelling states, which may be observed in the y-resolved local density-of-states (LDOS) shown in Figure S1 (Supporting Information).

The band structure of hNPG suggests that, in principle, positive charge carriers should exclusively propagate through cGNR states (setting $E - E_F$ within the valence band) and negative charge carriers through nGNR states (setting $E - E_F$ within the conduction band). To assess this, we use the GF method to simulate the local injection of charge from a model tip self-energy into hNPG devices electrostatically gated with either holes or electrons (see Methods for details). We systematically increase the gate both for holes and electrons from 0.1 to 1 e cell$^{-1}$, which approximately correspond to $0.36 \times 10^{13}$ and $3.6 \times 10^{13}$ e cm$^{-2}$, respectively, which are experimentally attainable (e.g. in graphene).[27] We note that such concentrations are experimentally attainable in graphene.[27] Gating leads to a narrowing of the energy spacing between the cGNR band onset and the nGNR band onset, both for valence and conduction bands (see Figure 1e; Figure S2, Supporting Information), thus making the hNPG look more like the "original" NPG. Such behavior may be understood as follows. Taking p-gating as an example, emptying the valence band leads to an effective positive charging of cGNRs, due to the primary cGNR character of the valence band. Such positive charging stabilizes cGNR states with respect to nGNR states, due to a lower coulomb repulsion, thus shifting the onset energies of the respective bands closer to each other. For n-gating the opposite phenomenon occurs: nGNRs are increasingly filled with electrons (due to the primarily nGNR character of the conduction band) which become negatively charged, increasing coulomb repulsion and thus their relative energy with respect to cGNRs. This narrowing of the gap between cGNR and nGNR bands reduces the range of gate voltages for which $E_F$ exclusively crosses either the cGNR band ($-0.89 \times 10^{13}$ e cm$^{-2}$) or the nGNR band ($0.89 \times 10^{13}$ e cm$^{-2}$), as shown in Figure S2 (Supporting Information). In Figure 1e we display the band structure for the p-gated case, which will be used to illustrate the transport properties of hNPG.

To evaluate quantum transport in hNPG, we take the $p_z$ elements of the DFT Hamiltonian to build an effective tight-binding (TB) model that captures the hNPG electronic structure under the effect of electrostatic gates (see Methods for details). This TB representation is then used to build a large-scale hNPG device composed of 291200 atoms, in which we simulate local injection of charge under a $-0.89 \times 10^{13}$ e cm$^{-2}$ p-gate (see associated band structure in Figure 1e). As seen in Figure 1f, local injection at the bottom region of the central cGNR leads to significant charge spreading, even upon setting $E_F$ to only cross the valence band with primarily cGNR character. This demonstrates that the band staggering of hNPG is not sufficient to confine injected currents to a single GNR, which we associate to the presence of tunneling states (Figure 1c; Figure S1, Supporting Information). We see, however, that the resulting Talbot interference, typical for NPGs,[23] displays an alternating pattern; that is, the bond transmission amplitude exclusively emerges in the cGNRs with nearly no amplitude in the nGNRs. This is a unique feature of hNPG which originates from the primarily cGNR character of the valence band that is electrically probed (see Figure 1e). If we locally contact an nGNR, we observe a significant reduction of the overall transmission amplitude (Figure S3c, Supporting Information) which may be associated with the smaller contribution of nGNR states at $E_F$ under the applied $-0.89 \times 10^{13}$ e cm$^{-2}$ p-gate (see associated LDOS in Figure S4, Supporting Information). Exactly the opposite scenario is obtained upon applying the corresponding n-gate ($+0.89 \times 10^{13}$ e cm$^{-2}$), which populates the nGNR conduction band (Figure S5a, Supporting Information): namely, the emergence of a nGNR-hosted Talbot interference pattern upon locally contacting a nGNR (Figure S5c, Supporting Information) and the near absence of transmission amplitude upon contacting a cGNR (Figure S5b, Supporting Information). Further increasing the gate voltage (either positive or negative) simultaneously populates both cGNR and nGNR bands, leading to a superposition of two Talbot interference patterns that emerge, with equal intensity, regardless of the contacted GNR (see Figures S6 and S7 (Supporting Information) associated with $-3.58 \cdot 10^{13}$ e cm$^{-2}$ and $+3.58 \cdot 10^{13}$ e cm$^{-2}$ gates, respectively). These two superimposed Talbot patterns are associated with the electronic wave functions of the two GNR subsystems (i.e. cGNRs and nGNRs; see below).

The results in Figure 1 clearly show that the band staggering of hNPG is insufficient to achieve confinement of currents in single GNRs. To better understand such complex transport behavior and propose directions to achieve current confinement, we propose a simple analytical model capturing the essential electronic features observed for hNPG (full details in Methods). This framework generalizes the Talbot interference model presented in ref. [23], based on the analogy between GNRs in NPGs and the channels within an array of laterally coupled optical waveguides. Here, the alternating sequence of nGNRs and cGNRs is represented by a periodic array of coupled waveguides, labeled $A_n$ and $B_n$, respectively, with $n$ denoting the unit cell index in the $y$ direction (**Figure** 2a). Each waveguide supports a single propagating mode with a local energy shift (propagation constant), $\varepsilon_A$ and $\varepsilon_B$. We assign $\varepsilon_A = -\alpha$ and $\varepsilon_B = \alpha > 0$, explicitly incorporating the band staggering observed in hNPG, and define $\kappa$ as the coupling constant between waveguides. In the optical analogy, $\alpha$ mimics a refractive index contrast, while $\kappa$ corresponds to the evanescent coupling between neighboring waveguides. We performed a DFT parametrization of $\kappa = 0.014$ Å$^{-1}$ and $\alpha = 0.065$ Å$^{-1}$. This parametrization describes the spatial features of the electron transmission of an n-gated hNPG ($0.36 \cdot 10^{13}$ e cm$^{-2}$) contacted in the bottom region of the central nGNR (see Note S1, Supporting Information). Below, we investigate the dependency of the interference described by the DDE model with respect to these two parameters.

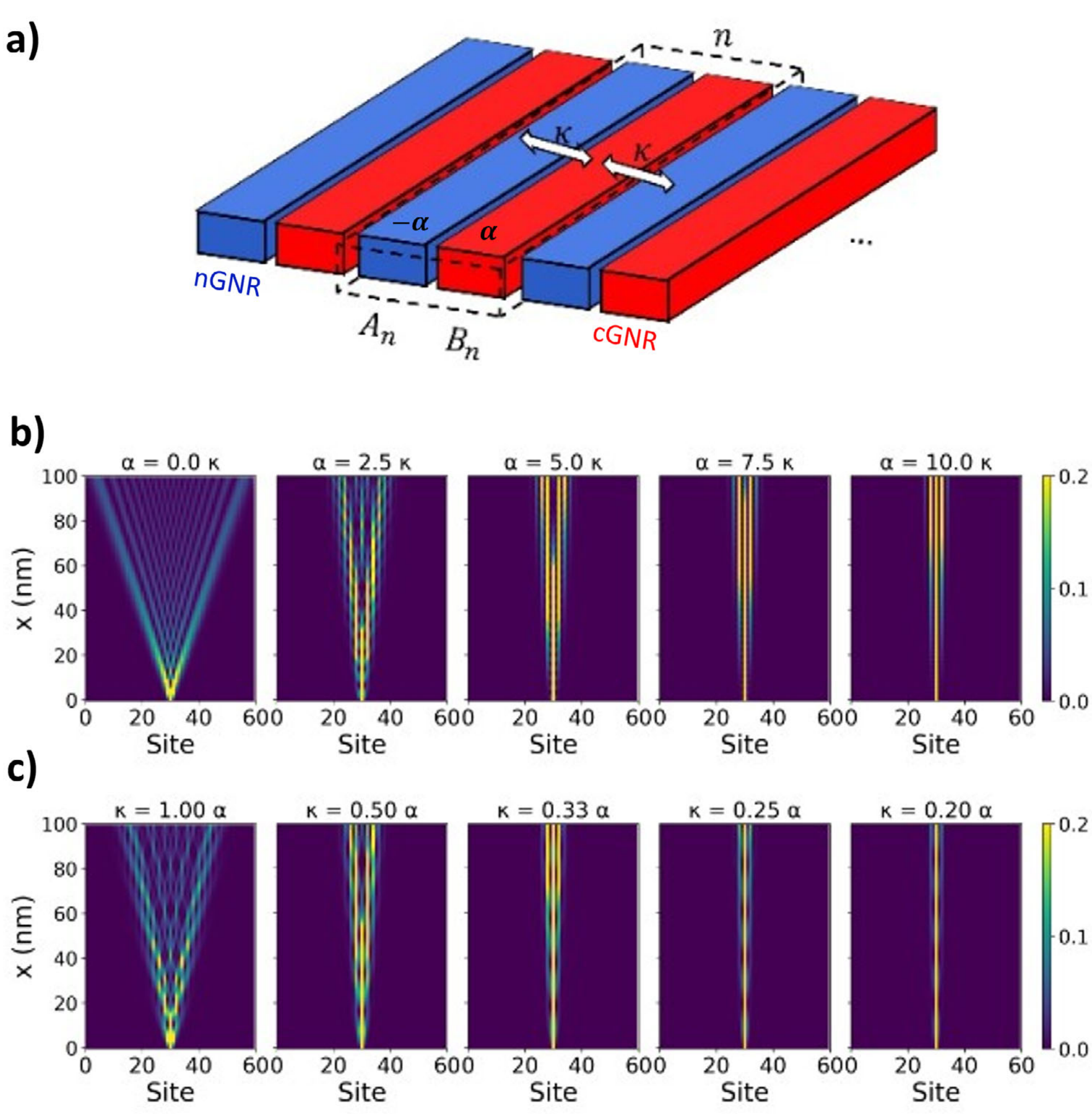

**Figure 2.** a) Schematic of the model representing hNPG as an infinite array of alternating $A_n$ ($B_n$) waveguides with potential propagation constant -$\alpha$ ($\alpha$), weakly coupled to their nearest neighbor by $\kappa$, with $n$ the unit cell index in the $y$-direction. b) Spatial propagation of waveguide amplitude obtained with the waveguide model for increasing values of the propagation constant ($\alpha$) at a fixed $\kappa = 0.014$ Å$^{-1}$. c) Spatial propagation of waveguide amplitude for decreasing coupling $\kappa$ at fixed $\alpha = 0.014$ Å$^{-1}$.

In Figure 2b, we show the propagating wave modes for increasing $\alpha/\kappa$, while fixing $\kappa = 0.014$ Å$^{-1}$ (see Note S1, Supporting Information for the rationale behind the choice of this value). The results exhibit a range of interference behavior, from the Talbot regime ($\alpha = 0$) to strong localization to $\sim$ 3 ribbons ($\alpha = 10\kappa$). At intermediate values of $\alpha/\kappa$, pronounced beating patterns in the cGNR waveguides ($B_n$) are observed. Strong (but not yet complete) localization of the field intensity along the contacted nGNR ($A_0$) is achieved only for large values of $\alpha \gg 10\kappa$, which may be challenging to achieve solely by chemical means. However, localization can also be achieved by combining the effects of moderate band staggering and reduced inter-ribbon coupling. This is demonstrated in Figure 2c, where the propagating modes are shown for increasing $\alpha/\kappa$, fixing $\alpha = 0.014$ Å$^{-1}$. For $\kappa = \alpha$, field intensities exhibit large spreading, albeit less pronounced than in NPG, due to the finite value of $\alpha$. In this intermediate regime, $\kappa \sim \alpha$, field intensities cannot be described by Talbot-interference.

As $\kappa$ decreases further, confinement is noticeably enhanced (see right panels in Figure 2c).

In light of these results, below we explore alternative hNPG designs to reduce the inter-ribbon coupling $\kappa$ and achieve fully confined charge propagation in the GNRs, which is of high interest for carbon nanocircuitry and, more generally, carbon nanoelectronics. Following prior proposals,[22] we first consider hNPG structures where every cGNR is surrounded by an increasing number of nGNRs, which should lead to a depletion of tunneling states and a reduced inter-cGNR coupling. More specifically, we consider hNPGs where cGNRs, instead of being separated by one nGNR as in hNPG (see Figure 1a), are separated by two nGNRs (hNPG-1,2, **Figure 3a**), three nGNRs (hNPG-1,3, Figure 3c), or four nGNRs (hNPG-1,4, Figure 3e). In this notation (hNPG-p,q) p and q are the number of cGNRs and nGNRs within the unit cell, respectively. In order to exploit the isolation of cGNR states to achieve 1D confinement, one needs to set $E_F$ to cross only the cGNR valence band. As done for hNPG, this requires a p-gate of approximately $-0.90 \cdot 10^{13}$ e cm$^{-2}$ (see Figure 1e), which we thus apply to hNPG-1,2, hNPG-1,3, and hNPG-1,4 during the optimization of the atomic and electronic structure (see Methods for details). To evaluate the effectiveness of these hNPG derivatives to confine currents, we simulate the injection of charge for the corresponding large-scale devices, locally probing the bottom region of the central cGNR. As we may see in Figure 3b, separating cGNRs by two nGNRs indeed leads to a lower degree of current spreading as compared to hNPG (Figure 1f), which is associated with an effective decrease of inter-cGNR coupling. By looking at the associated GNR-resolved fat band structure (Figure 3a), we see that hNPG-1,2 displays two nGNR bands near $E_F$ arising from the two nGNRs present in the hNPG-1,2 unit cell (see zoomed band structure). Such bands, by contributing to the y-resolved LDOS at such a doping level (Figure S8, Supporting Information), prevent a clean interrogation of cGNR states at $E_F$. Upon including more nGNRs in the unit cell, additional nGNR bands appear in the band structure nearby $E_F$ or even crossing it, as shown in Figure 3c,e for hNPG-1,3 and hNPG-1,4, respectively, and reflected in the associated LDOS (Figure S8, Supporting Information). Therefore, upon adding more nGNRs it gets increasingly difficult to set $E_F$ within the longitudinal regime of the cGNR valence band (that is, injecting carriers with a net momentum within the $\Gamma \rightarrow X$ segment) while preventing $E_F$ from crossing such emergent nGNR bands. As a consequence, despite the higher degree of confinement achieved by moving from hNPG (Figure 1f) to hNPG-1,4 (Figure 3f) due to an increased physical separation between cGNRs, there appear leaking currents due to such "new" nGNR bands near $E_F$. A possible manner to circumvent this would be to continually adjust the gate voltage as more nGNRs are added, such that $E_F$ only crosses the highest band associated to the cGNRs. However, this available energy window becomes progressively smaller as more nGNRs are added, as can be seen by comparing Figure 3a,c,e. If we assume transport at T = 300 K and let the threshold for closing this energy window be $k_B T$ (26 meV), then the window is already closed for hNPG-1,4 (Figure 3e). This effectively means that adding more than four nGNRs between each cGNR will not further improve confinement of charge carriers.

As a second strategy, we evaluate the use of QI in hNPGs. QI has been previously shown to be an effective route to engineer the transport properties of NPGs even under strong electrostatic[25] or dynamic (thermal) disorder.[26] To incorporate QI in the hNPG framework, we connect every GNR with its neighbors via phenyl rings in meta-configuration, giving rise to a new NPG whose geometry is shown in **Figure 4a**, which we label meta-hNPG. As seen in Figure 4b,c, the GNR contributions to the band structure in meta-hNPG are similar to those in hNPG (Figure 1d) – namely, the valence states are primarily localized in cGNRs and the conduction states in nGNRs (Figure 4c). However, the LDOS (Figure 4b) and GNR-resolved partial density of states (Figure S9, Supporting Information) highlight a significant difference. For hNPG, in the valence (conduction) band energy range, there is a non-negligible contribution by nGNR (cGNR) states (see arrows in Figure S9c, Supporting Information) associated with tunneling states (Figure S9b, Supporting Information). On the contrary, in meta-hNPG, such contributions are absent for both the valence and conduction bands (Figure S9f, Supporting Information), which we associate with the absence of tunneling states (Figure 4b). The suppression of tunneling states in this material upon gating is also confirmed (compare Figure S10 with Figure S4, Supporting Information).

As already mentioned, such tunneling states are instrumental in hNPGs to enable the coupling between GNRs of the same type and hence, ultimately, for the spreading of currents across GNRs. This relation seems to hold for meta-hNPG, displaying a negligible band dispersion in the $\Gamma \rightarrow Y$ path of the Brillouin zone (Figure 4c) and, simultaneously, a lack of tunneling states (Figure 4b). To evaluate the transport properties of meta-hNPG, we focus on the p-gated system under a gate of $-0.25$ e cell$^{-1}$ (as for hNPG) corresponding to $-0.72 \cdot 10^{13}$ e cm$^{-2}$. As may be seen in the corresponding band structure (Figure 4d) and LDOS (Figure S10, Supporting Information), this gate positions $E_F$ within the top valence band, primarily of cGNR character. By locally injecting currents in a large-scale meta-hNPG sample, shown in Figure 4e (270480 atoms), we see that full confinement of carriers is achieved within the single contacted cGNR up to at least 120 nm from the injection point (see small red dot at the bottom region of the sample). Similar to hNPG (Figure S3, Supporting Information), probing an nGNR gives rise to no transmission (Figure S11, Supporting Information). Increasing the applied gate ($-0.5$ e cell$^{-1}$) sets $E_F$ to cross both cGNR and nGNR bands (Figure S12a, Supporting Information) but, remarkably, full confinement of currents is still obtained regardless of the contacted GNR (Figure S12b,c, Supporting Information). Such level of collimation, well beyond both the hNPG (Figure 1f) and its lateral heterostructures (Figure 3), highlights the power of QI-engineering as a tool to control current flow in these nanomaterials.

The extremely efficient current confinement in meta-hNPG leads to the natural question of how it compares to bare QI-engineering strategies applied to NPGs without band staggering, such as in the so-called meta-NPG.[24] Meta-NPG has exactly the same geometry as meta-hNPG, but is composed entirely of cGNRs. Consequently, in meta-NPG, all GNRs are energetically degenerate, and hence current confinement solely originates from QI effects (i.e., due to meta-configured phenyl bridges). Given the strongly localized nature of transport in both cases, to discern possible differences, we simulate much longer devices (i.e., ≈45 × 610 nm$^2$, containing 710424 atoms), thus reaching the µm scale, and simulate the local injection of currents under a $-1.4 \cdot 10^{13}$ e cm$^{-2}$ p-gate ($-0.5$ e cell$^{-1}$). As shown in **Figure** 5a, current spreading sets in for meta-NPG after ≈100 nm from the injection point, monotonically increasing from that distance onward. On the contrary, injected currents in meta-hNPG stay fully collimated within the single contacted cGNR up to 600 nm (Figure 5b) and even 1 µm (Figure S13, Supporting Information), without showing any sign of spreading (see zoomed panels in Figure 5b; Figure S13a, Supporting Information). Additionally, at this p-gate value, such effective collimation works equally well upon contacting an nGNR (Figure S13b, Supporting Information). Given the width of 7-13-AGNRs (≈0.7 nm), our findings imply that one could send electric signals with subnanometer precision at least 1 µm away from the source using meta-hNPG-based carbon nanocircuits.

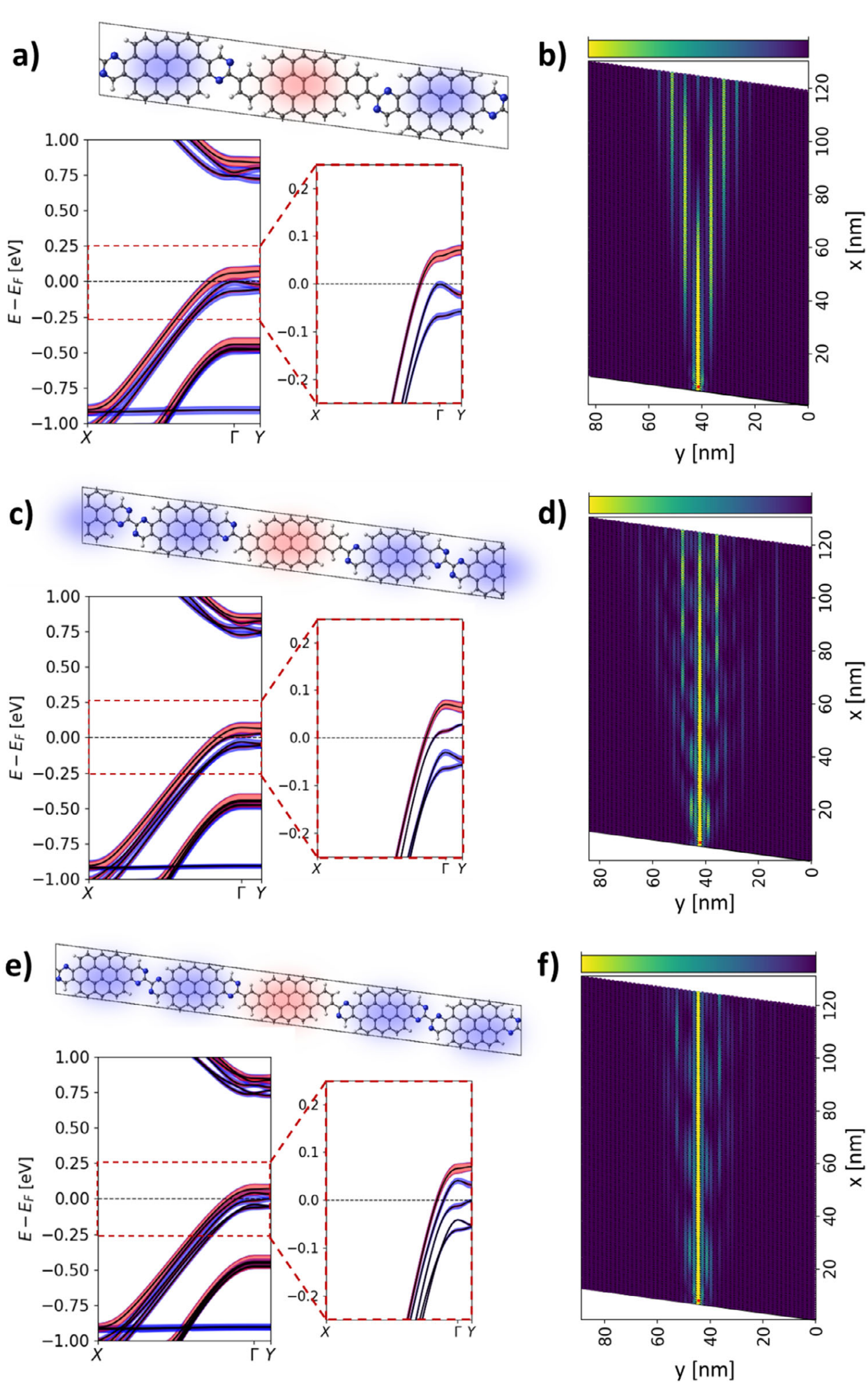

**Figure 3.** hNPG derivatives separating every cGNR, highlighted in red, by a) two nGNRs (hNPG-1,2), c) three nGNRs (hNPG-1,3), and e) four nGNRs (hNPG-1,4), highlighted in blue. All systems have been optimized under a p-gate of $-0.90 \cdot 10^{13}$ e $cm^{-2}$. Below each geometry the associated GNR-resolved fat band structure (cGNR: red; nGNR: blue) is shown, along with a zoomed view around $E_F$. Associated bond transmission maps of ≈85 × 122 $nm^2$ samples contacted in the bottom region of the central cGNR (see small red dot) are shown in b) hNPG-1,2 (285600 atoms), d) hNPG-1,3 (291200 atoms), and f) hNPG-1,4 (308000 atoms). All color bars in the bond transmission maps range from 0 to 0.045.

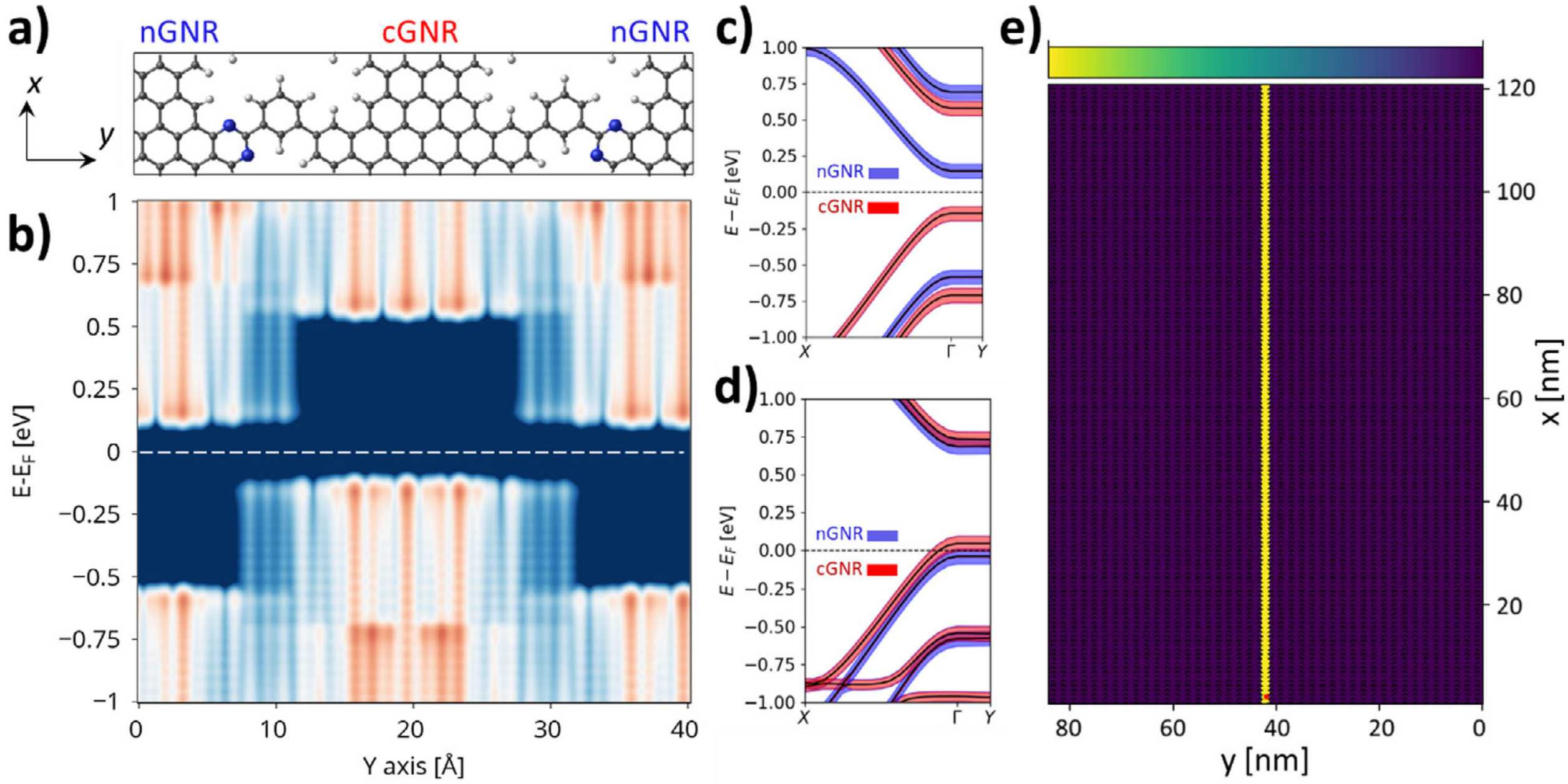

**Figure 4.** a) Optimized geometry of meta-hNPG and corresponding b) $y$-resolved LDOS. The colors are in a log scale, with dark red (blue) indicating a high (low) LDOS. $E_F$ is indicated with a dashed white line. Associated GNR-resolved fat band structure (cGNR: red; nGNR: blue) for c) the charge-neutral meta-NPG and d) the $-0.72 \cdot 10^{13}$ e cm$^{-2}$ p-gated material ($-0.25$ e cell$^{-1}$). e) Bond transmission map of an $\approx 85 \times 122$ nm$^2$ sample of the p-gated meta-hNPG upon locally contacting the central cGNR at the bottom region of the device (see small red dot). The color bar ranges from 0 to 0.045.

We have found that such high confinement behavior in meta-hNPG appears to be gate-dependent. For example, if we n-gate the material with +0.5 e cell$^{-1}$ (+1.44 $\cdot$ 10$^{13}$ e cm$^{-2}$), thus setting $E_F$ within the conduction band, contacting a cGNR leads to a slight confinement degradation after 300 nm and to appreciable spreading of currents to the nearest and next-nearest neighbor cGNRs for distances ≈1 µm away from the source point (Figure S14a, Supporting Information; note that such spreading occurs exclusively throughout cGNRs, as for hNPG). Curiously, contacting an nGNR at this same gate (+1.44 $\cdot$ 10$^{13}$ e cm$^{-2}$) displays perfect current collimation up to one µm (Figure S14b, Supporting Information), which could be attributed to the electron–hole asymmetry in this material due to the presence of N atoms. In any case, the overall confinement in such n-gated meta-hNPG is still well beyond that of meta-NPG under the same conditions, where a noticeable spreading of currents begins as close as 60 nm from the point of injection (Figure S14c, Supporting Information).

## 3. Conclusion

In this work, we have studied the transport properties of hNPG systems, which are 2D arrays of alternating cGNRs and nGNRs. Transport simulations of hNPG have shown that injected currents significantly spread throughout the material, leading to the well-known Talbot interference pattern as reported in other NPGs.[23,24] However, the band staggering in hNPG leads to current spreading through the cGNR sub-system, the nGNR subsystem, or both, depending on the applied gate and the contacted GNR. Such unique effects may be fully captured with an analytical model that exclusively considers two key parameters: i) the on-site energy difference between GNRs of different types (propagation constant, $\alpha$) and ii) the effective coupling between neighbouring GNRs ($\kappa$). This provides a physical understanding of the observed interference phenomena in the staggered system. The model was parameterized and validated by linking $\alpha$ and $\kappa$ to the hNPG electronic band structure from DFT, and reproducing the bond transmission maps obtained using the TB-GF approach. The model highlights that Talbot interference is hosted in every other ribbon of hNPG, featuring a beating pattern that depends solely on $\alpha$, as discussed in the Supporting Information. The model also suggested that current confinement can be enhanced by decreasing the coupling constant $\kappa$ or by significantly increasing the band staggering $\alpha$. The latter could be achieved, for instance, by adsorption of functional groups near the pores or by substitution of N dopants with more electronegative species. However, without a concurrent decrease in $\kappa$, it would still not allow for full confinement.

Therefore, in the second part of the work, we focused on evaluating possible alternative hNPG design strategies to reduce the inter-ribbon coupling. First, we considered increasing the number of nGNRs between cGNRs. However, this strategy leads to the emergence of new nGNR bands very close to, or even crossing, the Fermi level in the gated systems. This leads to parasitic current spreading, preventing full confinement of charge carriers in single GNRs, which is needed for nanoelectronic applications requiring nearly atomic precision. Nevertheless, we have presented an alternative strategy that solves these issues. This is based on embedding QI-effects in the hNPG by connecting every GNR with meta-configured phenyl ring bridges. In the resulting material, which we label meta-hNPG, tunnelling states are nearly absent, which leads to full confinement of injected currents within a single GNR. Further, we find that meta-hNPG represents a significant improvement in terms of current collimation over its predecessor, the deeply studied meta-NPG. This highlights a synergistic effect present in meta-hNPG, where a combination of the band-staggering profile and QI gives rise to a level of control over current flow that is not possible otherwise. Despite a certain dependence on the applied gate (p-gate vs n-gate) and the contacted GNR (cGNR vs nGNR), meta-hNPG allows one to send electric signals with sub-nanometer precision for distances up to at least one micrometer away from the point of injection.

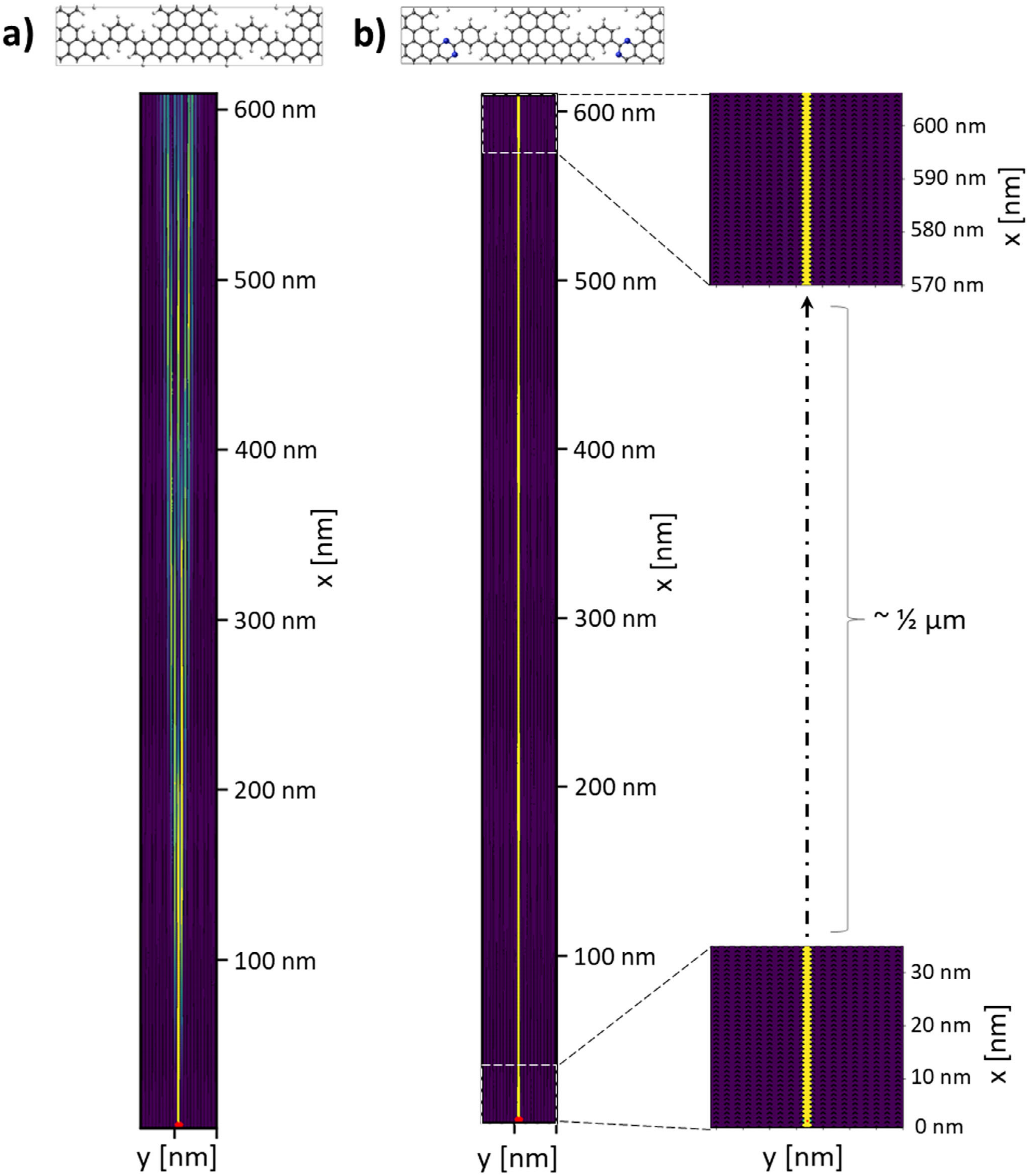

**Figure 5.** Bond transmission maps of $-1.44 \cdot 10^{13}$ e cm$^{-2}$ p-gated a) meta-NPG and b) meta-hNPG samples, with a size of $45 \times 610$ nm$^2$. In both cases, charge is injected locally in the central cGNR at the bottom region of the device (see small red dot). The corresponding optimized atomic geometries are provided in the top panels. For meta-NPG (a) the corresponding band structure is shown in Figure S14c (Supporting Information). For the meta-hNPG (b) zoomed bond transmission maps are provided for the bottom and top regions of the device. All color bars in the bond transmission maps range from 0 to 0.045.

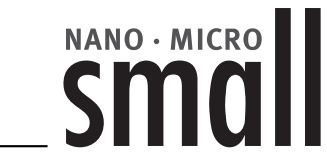

It is worth highlighting that this level of electron confinement is unprecedented in 2D materials. While multiple studies have reported transport confinement via lithographically fabricated gate-defined quantum point contacts[28] or parabolic p-n junctions,[29,30] the level of achieved collimation typically goes down to a few hundreds of nanometers, only having recently got below 10 nm in wrinkled 2D layers.[31] Alternatively, our molecular-engineering approach allows designing 0.7 nm-wide channels with full collimation up to a micron away from the source, which is a qualitative step beyond the aforementioned approaches. As for experimental feasibility, the tremendous recent progress in the on-surface synthesis of different types of NPGs[10–13,22] and their integration in solid-state devices[8,19,32] highlights that the realization of our novel NPG designs and their technological implementation is within reach. Finally, while some works have shown that the Talbot interference pattern in NPGs is susceptible to local scatterers,[23] more recent studies show its resilience to a high density of atomic vacancies.[33] Additionally, QI has been predicted to be robust against different disorder sources such as electron–electron and electron-phonon scattering.[34] Along this line, it has recently been found that QI-engineering of transport in NPGs survives both under electrostatic and thermal (at 300 K) disorder.[25,26] Thus, all in all, our findings for meta-hNPG should be robust under experimentally realistic conditions, making this material a very promising platform for future carbon nanocircuitry and, more generally, carbon nanoelectronics.

## 4. Experimental Section

*Density Functional Theory*: The atomic and electronic structure of hNPGs has been optimized with density functional theory (DFT) using the PBE (GGA) exchange-correlation functional,[35] a single-$\zeta$ basis set with 0.02 Ry energy shift, norm-conserving Troullier–Martins pseudopotentials, a real-space mesh cutoff of 400 Ry, and a 51 × 15 × 1 Monkhorst–Pack k-point mesh. The SIESTA software[36] has been used for all DFT calculations. Periodic geometries have been optimized until all forces were below 0.01 eV Å$^{-1}$, also minimizing cell parameters except the c-parameter (*z*-direction) which was kept fixed to avoid inter-layer interactions (50 Å). The external application of electrostatic gates was simulated by placing fixed charges on a plane parallel to the 2D material positioned 10 Å away. For every applied gate, the atomic structure and cell parameters of the system were fully optimized to account for any relaxation mechanisms caused by the addition or removal of charge. All post analysis, including fat-band structures, LDOS, PDOS and band structures, was done with SISL.[37]

*Quantum Transport Simulations*: Transport simulations were performed using a well-established parameterization approach for $\pi$-conjugated 2D materials,[13,23,24,38–42] by extracting the onsite and coupling parameters of $p_z$ orbitals for all C and N atoms from the converged DFT Hamiltonians and overlap matrices. These parameters were used to parameterize computationally efficient tight binding (TB) models generated with the open-source SISL Python package.[37] In particular, all DFT Hamiltonian elements involving $p_z$ atomic orbitals of C and N atoms were exploited to construct the TB Hamiltonian. The resulting TB band structure was slightly misaligned with respect to the DFT one. To solve this, all onsite energies in the TB Hamiltonian were rigidly shifted to match the DFT mid-gap level of the neutral systems, the valence band edge of the p-gated systems or the conduction band edge of the n-gated systems. This reproduced the DFT band structure within the energy range of interest for the quantum transport simulations (see Figure S15a, Supporting Information), while retaining the full interaction range of the DFT basis set. Consequently, such lighter TB models capture the impact of sub-nanometer quantum dipoles, in-gap tunneling states, electrostatic gates, and hydrogen passivation on the electronic structure. Including other orbitals from DFT into the TB Hamiltonian, such as N-$d_{xz}$ and N-$d_{yz}$, slightly improves the match with DFT bands (see Figure S15b, Supporting Information) at the cost of reducing the computational efficiency and the maximum achievable device size. Large-scale devices were constructed by replicating the unit cell in both in-plane directions until reaching the desired system sizes, being indicated in the main text along with the corresponding total number of atoms. Quantum transport in these devices was simulated using the TBtrans code,[43] which was based on an efficient implementation of the Green's function formalism. Current injection at specific sites of the device was simulated via a constant, on-site broadening self-energy in the device Green's function,[24] effectively modelling a probe microscopy tip in chemical contact with one atom. Current was drained by in-plane semi-infinite hNPG electrodes along the *x*-direction (the GNR direction). Complex absorbing potential walls were applied along the *y*-direction, to avoid electronic back-reflection at the nonperiodic boundaries of each device.[29,44] Bond transmission maps were generated using SISL to illustrate the spatial distribution of injected currents,[29] using a color scale proportional to the transmission magnitude at each atom, with high transmission displayed in bright yellow and low to zero transmission in dark purple (all color bars in the article span the same range of values, 0 to 0.045). Bond transmissions were computed by TBtrans at the Fermi level only. In terms of performance, the largest TBtrans device simulation presented in this work, consisting of a 1017 × 36 nm$^2$ meta-hNPG system with nearly $10^6$ atoms (968760; 1 $p_z$ orbital/atom), was performed using the OpenMP-parallelized TBtrans program (version 5.1-MaX-50-g4df4b742f) on a 2 × 10-core Intel Xeon E5-2660v3 (2.6GHz) node with 128 GB RAM, considering 20 OpenMP threads. It requires ≈55 GB of RAM (memory requirements represent the main constraint for this type of simulation) and ≈10 min of CPU time. Set-up and post-processing (i.e., plotting bond transmission maps) in SISL for this simulation required ≈5 and ≈35 min, respectively.

*Analytical Model*: The wave equation for the coupled modes, proposed to describe electron transport in hNPG, corresponds to the system of coupled discrete differential equations

$$\begin{cases} i\frac{dA_n}{dx} + \kappa\ (B_n + B_{n-1}) = \epsilon_A\ A_n, \\ i\frac{dB_n}{dx} + \kappa\ (A_n + A_{n+1}) = \epsilon_B\ B_n, \end{cases} \tag{1}$$

where $A_n$ ($B_n$) waveguides within the nth unit cell in the y direction were associated with nGNRs (cGNRs) with a local energy shift $\varepsilon_A$ ($\varepsilon_B$) and $\kappa$ as the nearest neighbor coupling. The weak-coupling regime was considered, with only nearest-neighbor coupling between waveguides. Here, we investigate single ribbon injection, corresponding to a delta source at a lateral waveguide (for instance $A_0$) at a fixed position $\delta(x - x_0)$. Solutions $Y_n \in \{A_n, B_n\}$ can be expressed as $\boldsymbol{Y}_n\ (x) = \int_{-\pi}^{\pi} dk e^{ink} \boldsymbol{\gamma}_k(x)$, where $k$ is linear momentum. Fourier transforming into $\omega$-space allows us to rewrite the coupled differential equations in an algebraic form:

$$(-\omega I + \boldsymbol{\Omega})\boldsymbol{\gamma}_k\ (\omega) = \begin{pmatrix} 1 \\ 0 \end{pmatrix} \tag{2}$$

where $\boldsymbol{I}$ is a $2 \times 2$ identity matrix, $\boldsymbol{\Omega} = \begin{pmatrix} \alpha & 2t^* \cos(k/2) \\ 2t \cos(k/2) & -\alpha \end{pmatrix}$, $\boldsymbol{\gamma}\ (\omega) = \begin{pmatrix} a_k(\omega) \\ b_k(\omega) \end{pmatrix}$ and $t = \kappa e^{ik/2}$being the k-dependent complex coupling.

Real-space solutions can be obtained by Fourier transformation back from $(k, \omega)$ into $(n, x)$:

$$\begin{cases} A_n\ (x) = -i \int_{-\pi}^{\pi} dk\, e^{ink} \cos(\omega_0 x) - i \int_{-\pi}^{\pi} dk e^{ink} \frac{\alpha}{\omega_0} \sin(\omega_0 x) \\ B_n\ (x) = \int_{-\pi}^{\pi} dk\, e^{ink} 2t \cos(k/2)\ \sin(\omega_0 x) \end{cases} \tag{3}$$

with $\omega_0^2 = \alpha^2 + |2t \cos(k/2)|^2$.

In this work, focus was placed on two limiting cases. For $\alpha \ll \kappa$, solutions reduce to:

$$\begin{cases} A_n\ (x) \approx -i J_n\ (2\kappa x) \\ B_n\ (x) \approx J_{n-1}\ (2\kappa x) - J_{n+1}\ (2\kappa x) \end{cases} \tag{4}$$

which recovers the Talbot interference pattern of NPG.[23] Here $J_n$ denotes the Bessel functions of the first kind. In the opposite regime, with $\alpha \gg \kappa$, solutions become:

$$\begin{cases} A_n\ (x) \approx -i J_n \left( \frac{2\kappa^2}{\alpha} x \right) e^{-i\alpha x} \\ B_n\ (x) \approx \frac{2\kappa}{\alpha} \sin(\alpha x)\ J_n \left( \frac{2\kappa^2}{\alpha} x \right) e^{-i\alpha x} \end{cases} \tag{5}$$

In this limit, Talbot-like interference is hosted only by $A_n$ sub-system, with wavenumber rescaled by a factor $\kappa/\alpha \ll 1$, while waveguide modes intensity on $B_n$ is suppressed accordingly. Periodic modulation of the field waveguide modes in $B_n$ originates from the potential (propagation constant) mismatch.

## Supporting Information

Supporting Information is available from the Wiley Online Library or from the author.

## Acknowledgements

G.C. and I.A. contributed equally to this work. G.C. acknowledges the project SAMOTHRACE “Sicilian micro and nanotechnology research and innovation centre”, founded by PNRR-MUR (ECS_00000022, CUP B63C22000620005) for partial support. I.A. is grateful for a Juan de la Cierva postdoctoral grant (FJC2019-038971-I) from the Ministerio de Ciencia e Innovación. ICN2 is funded by the CERCA Programme from Generalitat de Catalunya, has been supported by the Severo Ochoa Centres of Excellence programme [SEV-2017-0706], and is currently supported by the Severo Ochoa Centres of Excellence programme, Grant CEX2021-001214-S, both funded by MCIN/AEI/10.13039.501100011033. A.E.A.M. was funded by Independent Research Fund Denmark, Grant 0.46540/3103-00229B. M.P. acknowledges support from PID2022-139776NB-C61, funded by Spanish MCIN/AEI https://doi.org/10.13039/501100011033, by the ERDF, A way of making Europe, and by Generalitat de Catalunya (Grant 2021SGR01519). This work was also supported by MICIN with European funds (NextGenerationEU; PRTR-C17.I1) and by 2021 SGR 00997 funded by Generalitat de Catalunya.

## Conflict of Interest

The authors declare no conflict of interest.

## Data Availability Statement

The data that support the findings of this study are available from the corresponding author upon reasonable request.

## Keywords

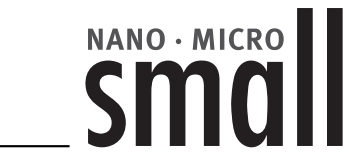

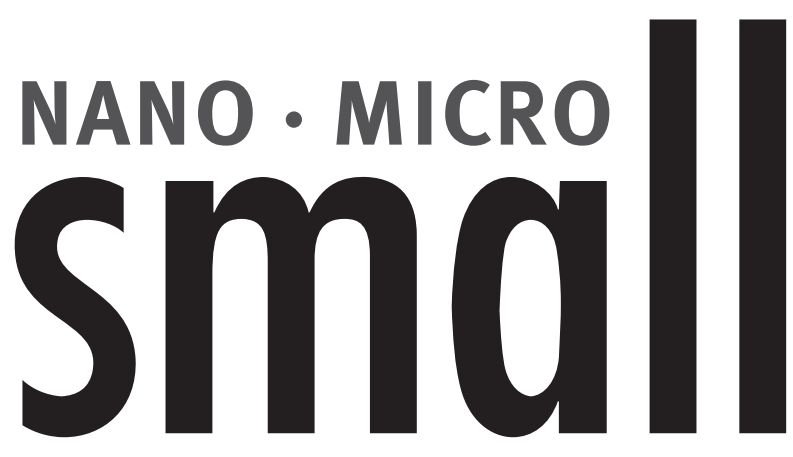

Supporting Information

Quantum Transport in Nitrogen-Doped Nanoporous Graphenes

*Gaetano Calogero, Isaac Alcón*, Alan E. Anaya Morales, Nick Papior, Pol Febrer, Aron W. Cummings, Miguel Pruneda, Stephan Roche and Mads Brandbyge*

# Supporting Information

## Quantum transport in nitrogen-doped nanoporous graphenes

*Gaetano Calogero, Isaac Alcón*, Alan Morales, Nick Papior, Pol Febrer, Aron W. Cummings, Miguel Pruneda, Stephan Roche and Mads Brandbyge*

# Supporting Information

## Quantum transport in nitrogen-doped nanoporous graphenes

Gaetano Calogero[1†], Isaac Alcón[2,3†*], Alan Morales[4], Nick Papior[5], Pol Febrer[2], Aron W. Cummings[2], Miguel Pruneda[2,6], Stephan Roche[2,7] and Mads Brandbyge[4]

[1]*National Research Council, Institute for Microelectronics and Microsystems (CNR-IMM), Zona Industriale, Strada VIII, 5, 95121 Catania, Italy*

[2]*Catalan Institute of Nanoscience and Nanotechnology (ICN2), CSIC and BIST, Campus UAB, Bellaterra, 08193 Barcelona, Spain*

[3]*Institute of Theoretical and Computational Chemistry, Department of Materials Science and Physical Chemistry, Universitat de Barcelona, C/ de Martí i Franquès, 1-11, Les Corts, 08028, Barcelona, Spain*

[4]*Department of Physics, Technical University of Denmark, DK-2800 Kongens Lyngby, Denmark*

[5]*Computing Center, Technical University of Denmark, DK-2800 Kongens Lyngby, Denmark*

[6]*Nanomaterials and Nanotechnology Research Centre (CINN), CSIC, Avenida de la Vega 4-6. 33940, El Entrego, Spain*

[7]*ICREA, Institució Catalana de Recerca i Estudis Avançats, 08070 Barcelona, Spain*

[†]equally contributed

*Corresponding author: ialcon@ub.edu

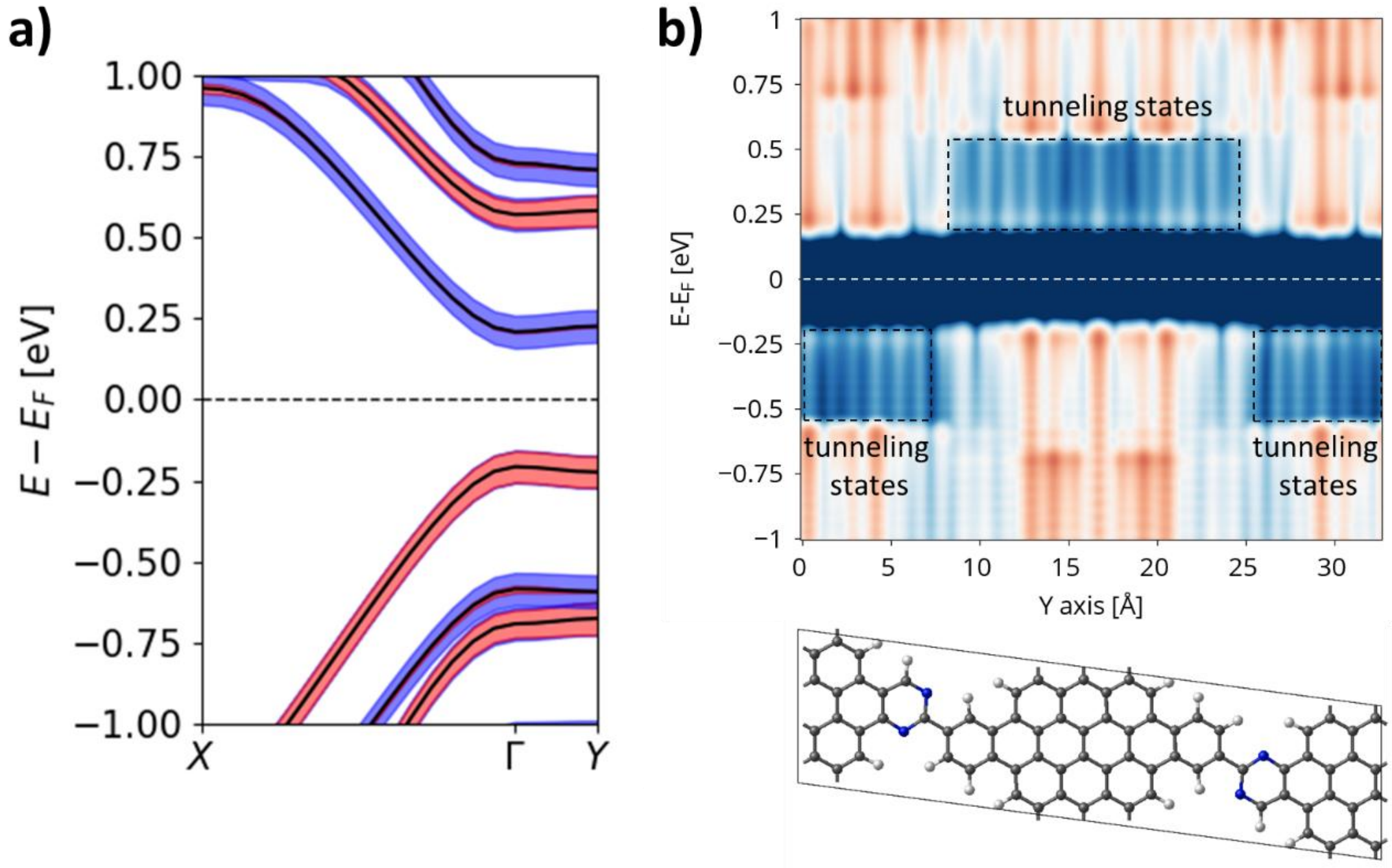

Figure S1. a) GNR-resolved fat band structure (cGNR: red; nGNR: blue) of hNPG in the neutral state. b) Local density of states (LDOS) along the y-axis, showing that valence (conduction) states are mainly located in the cGNR (nGNR), and that tunnelling states exist at all energies beyond the band edges. The colors are in a log scale, with dark red (blue) indicating a high (low) LDOS. The Fermi level ($E_F$) is indicated with a dashed white line. The hNPG atomic structure is provided in the bottom panel for a one-to-one comparison with the LDOS plot.

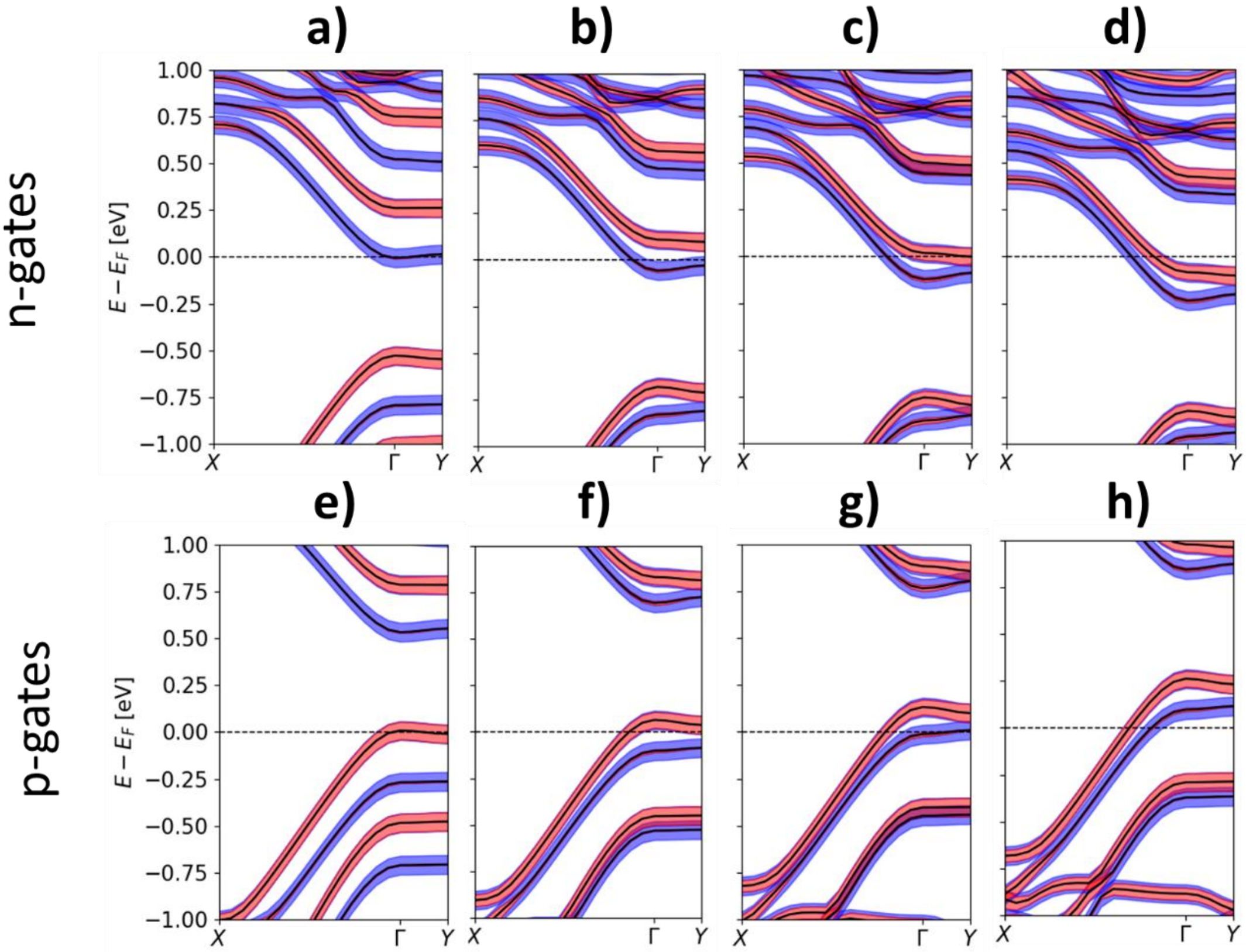

Figure S2 GNR-resolved fat band structures (cGNR: red; nGNR: blue) of hNPG for n-gates a) +0.1 e/cell (+0.36 · $10^{13}$ e/cm$^2$), b) +0.25 e/cell (+0.89 · $10^{13}$ e/cm$^2$), c) +0.5 e/cell (+1.79 · $10^{13}$ e/cm$^2$), d) +1.0 e/cell (+3.58 · $10^{13}$ e/cm$^2$), and p-gates e) -0.1 e/cell (-0.36 · $10^{13}$ e/cm$^2$), f) -0.25 e/cell (-0.89 · $10^{13}$ e/cm$^2$), g) -0.5 e/cell (-1.79 · $10^{13}$ e/cm$^2$), h) -1.0 e/cell (-3.58 · $10^{13}$ e/cm$^2$).

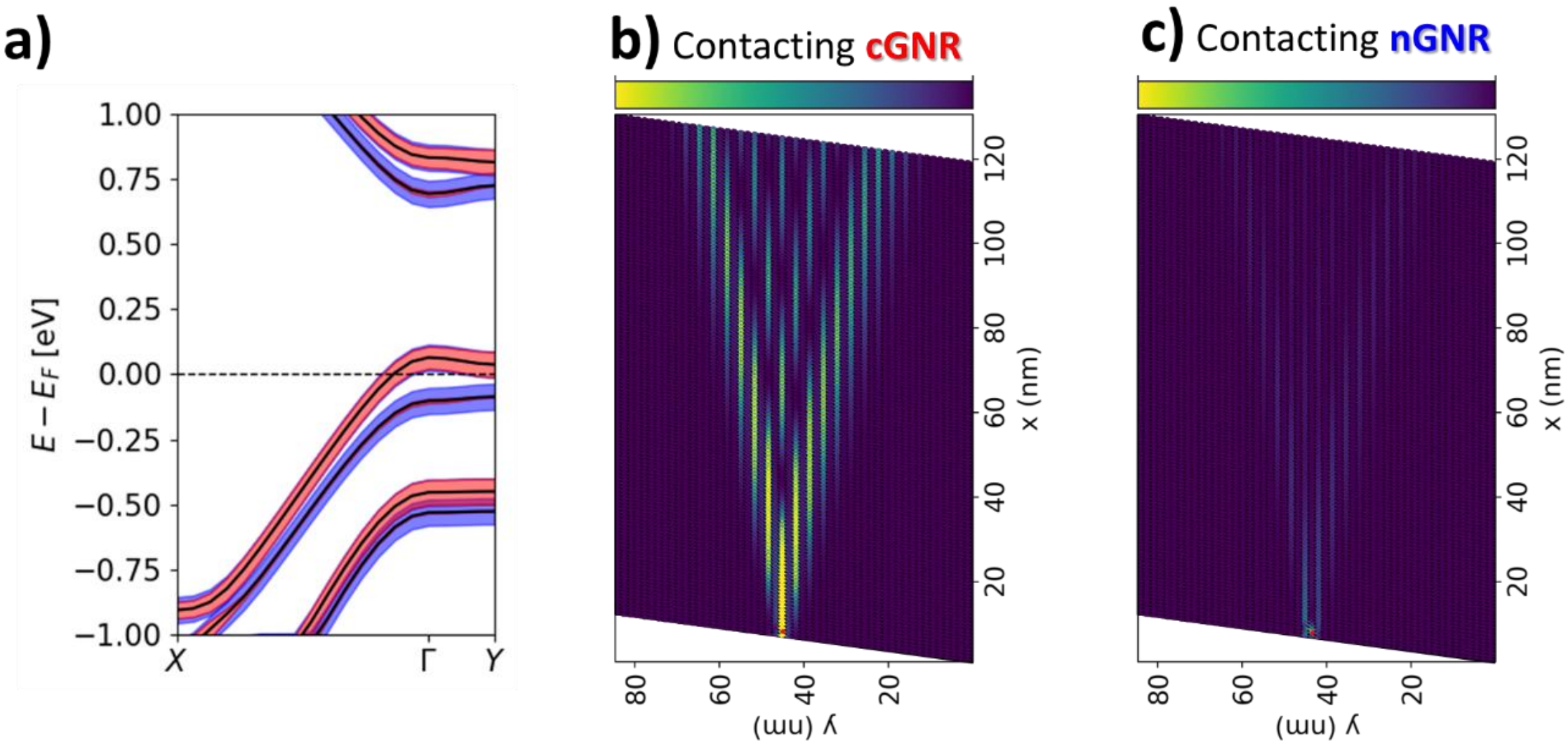

Figure S3. a) GNR-resolved fat band structure (cGNR: red; nGNR: blue) of hNPG for -0.25 e/cell gating (-0.89 · $10^{13}$ e/cm$^2$) and corresponding bond transmission maps of a ~85 x 122 nm$^2$ sample upon locally contacting b) a cGNR or c) a nGNR at the bottom part of the sample (see small red dot). The color bars in the bond transmission maps range from 0 to 0.045. The Hamiltonian of these large-scale samples has been obtained by taking $p_z$ orbitals from the DFT Hamiltonian (see Methods for details).

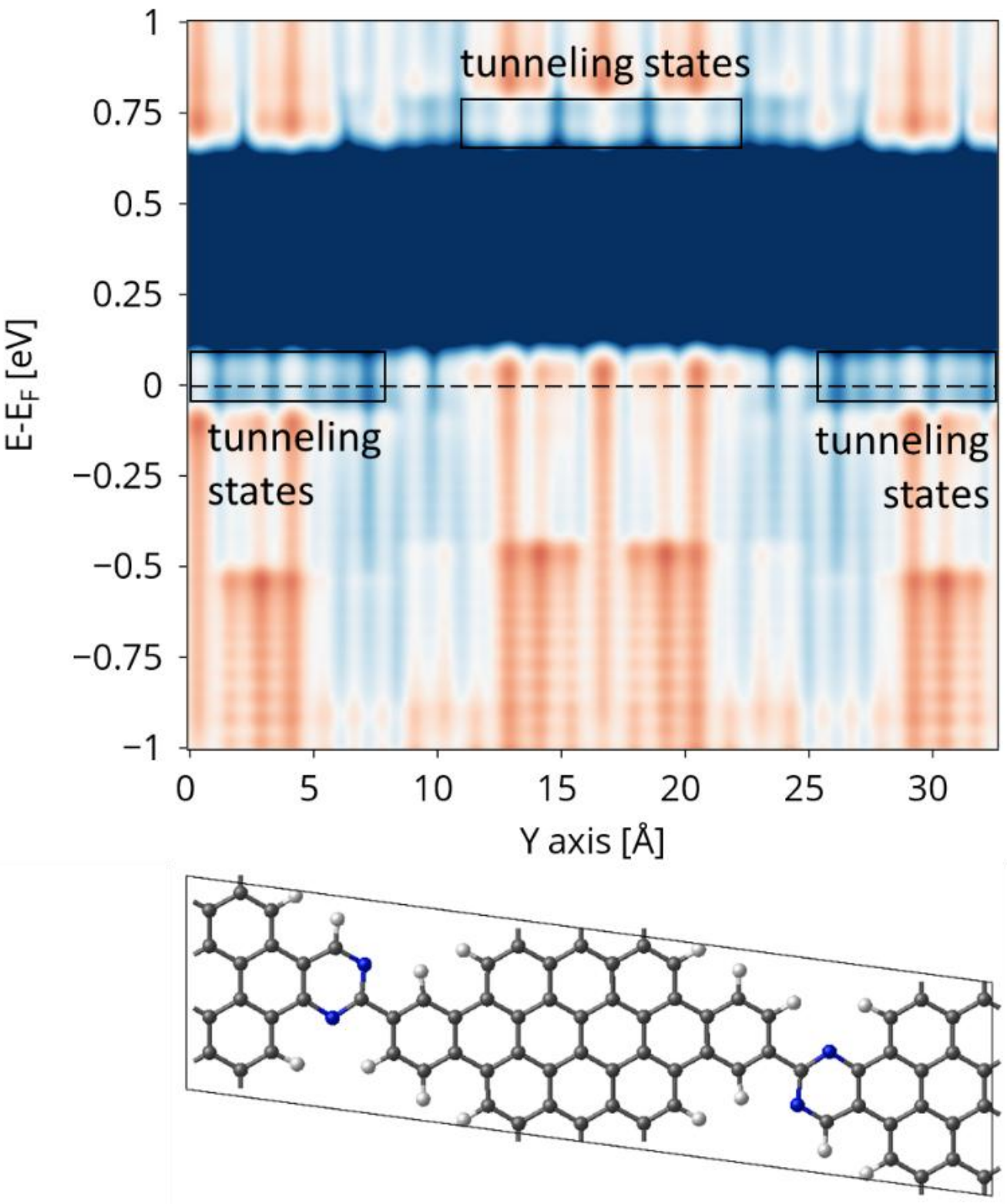

Figure S4. LDOS along the y-axis for the -0.25 e/cell gated hNPG (-0.89 · $10^{13}$ e/cm$^2$). The colors are in a log scale, with dark red (blue) indicating a high (low) LDOS. $E_F$ is indicated with a dashed black line. The corresponding band structure is shown in Figure S3a. The hNPG atomic structure is provided in the bottom panel for a one-to-one comparison with the LDOS plot.

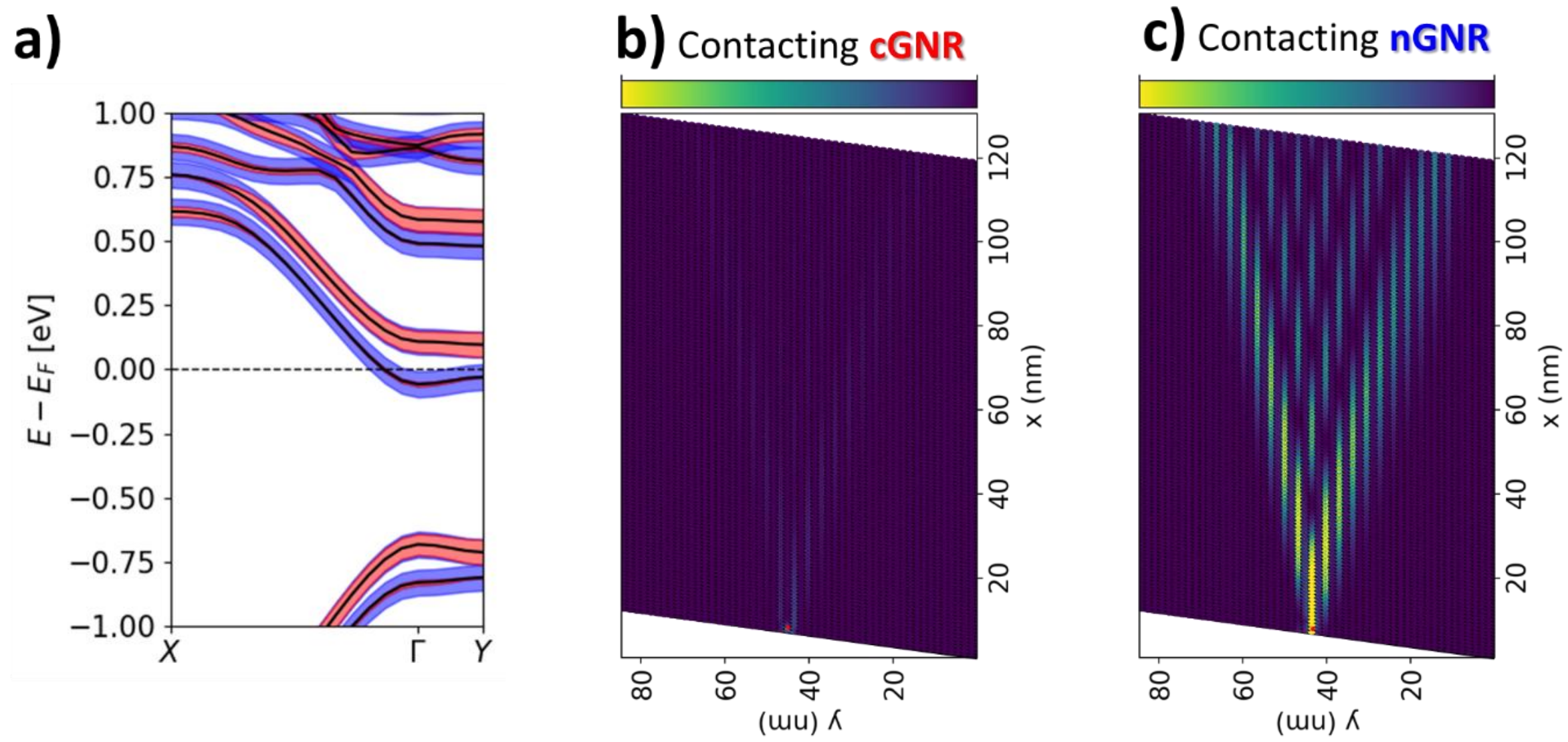

Figure S5. a) GNR-resolved fat band structure (cGNR: red; nGNR: blue) of hNPG for +0.25 e/cell gating (+0.89 · $10^{13}$ e/cm$^2$) and corresponding bond transmission maps of a ~85 x 122 nm$^2$ sample upon locally contacting b) a cGNR or c) a nGNR at the bottom part of the sample (see small red dot). All color bars in the bond transmission maps range from 0 to 0.045. The Hamiltonian of these large-scale samples has been obtained by taking $p_z$ orbitals from the DFT Hamiltonian (see Methods for details).

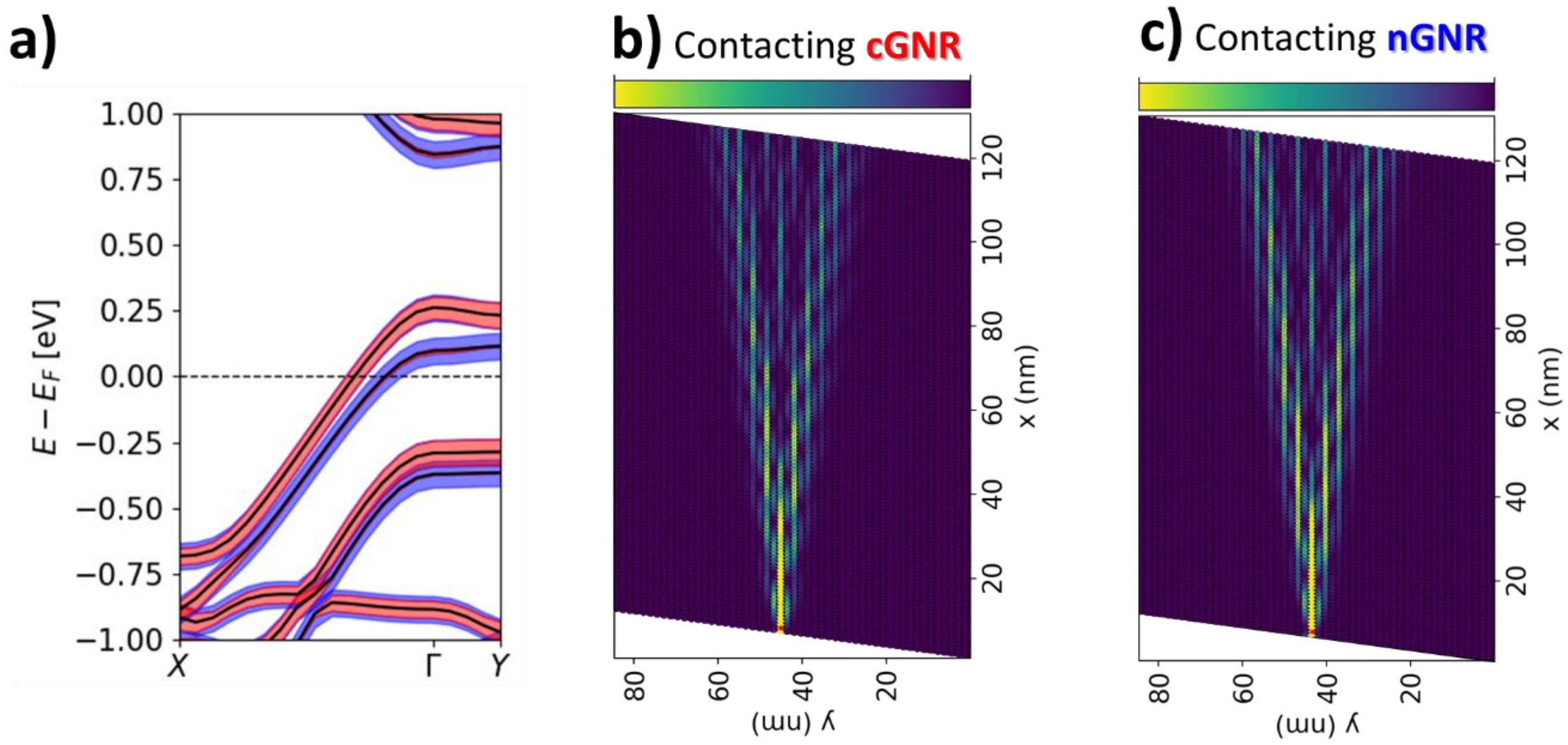

Figure S6. a) GNR-resolved fat band structure (cGNR: red; nGNR: blue) of hNPG for -1.0 e/cell gating (-3.58 · $10^{13}$ e/cm$^2$) and corresponding bond transmission maps of a ~85 x 122 nm$^2$ sample upon locally contacting b) a cGNR or c) a nGNR at the bottom part of the sample (see small red dot). All color bars in the bond transmission maps range from 0 to 0.045. The Hamiltonian of these large-scale samples has been obtained by taking $p_z$ orbitals from the DFT Hamiltonian (see Methods for details).

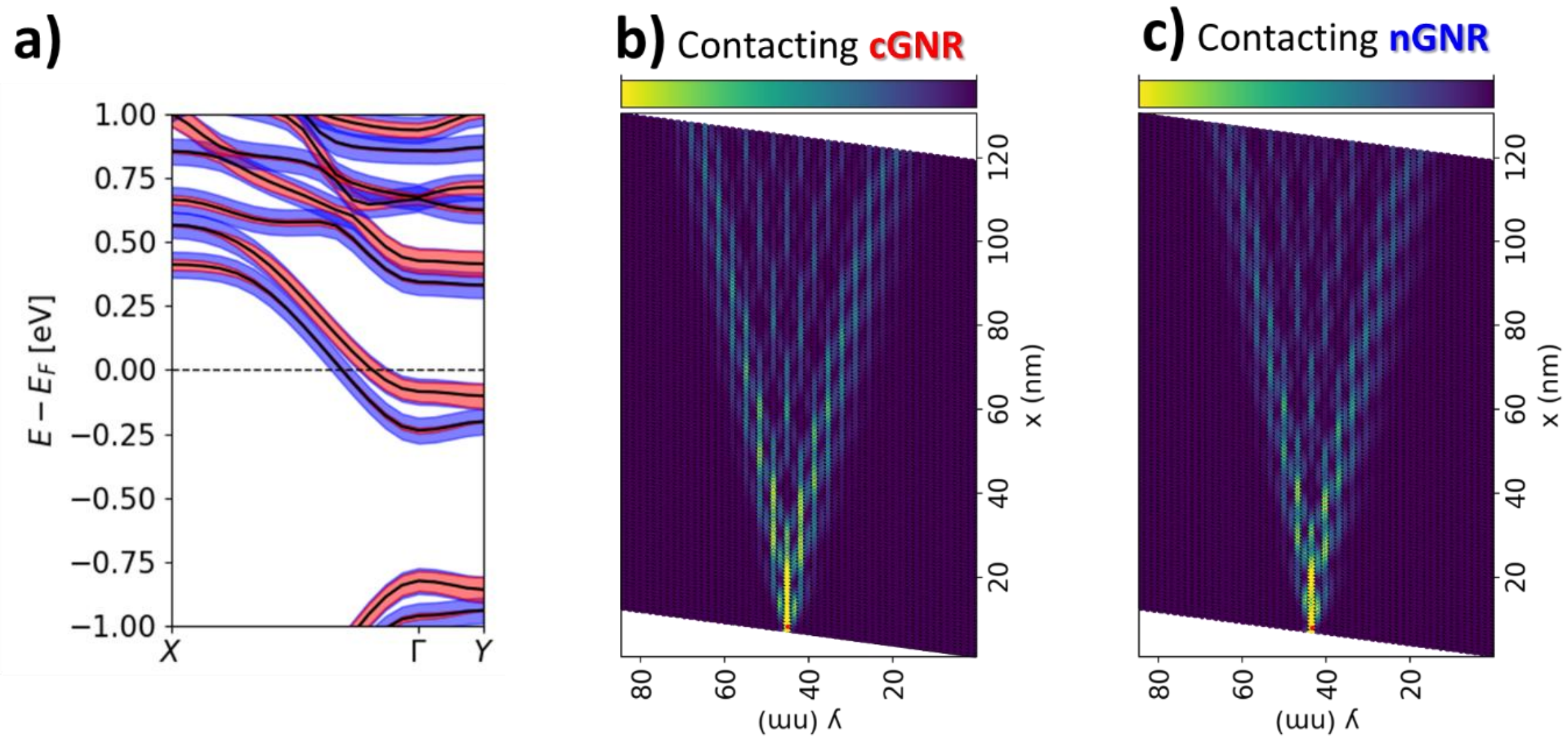

Figure S7. a) GNR-resolved fat band structure (cGNR: red; nGNR: blue) of hNPG for +1.0 e/cell gating (+3.58 · $10^{13}$ e/cm$^2$) and corresponding bond transmission maps of a ~85 x 122 nm$^2$ sample upon locally contacting b) a cGNR or c) a nGNR at the bottom part of the sample (see small red dot). All color bars in the bond transmission maps range from 0 to 0.045. The Hamiltonian of these large-scale samples has been obtained by taking $p_z$ orbitals from the DFT Hamiltonian (see Methods for details).

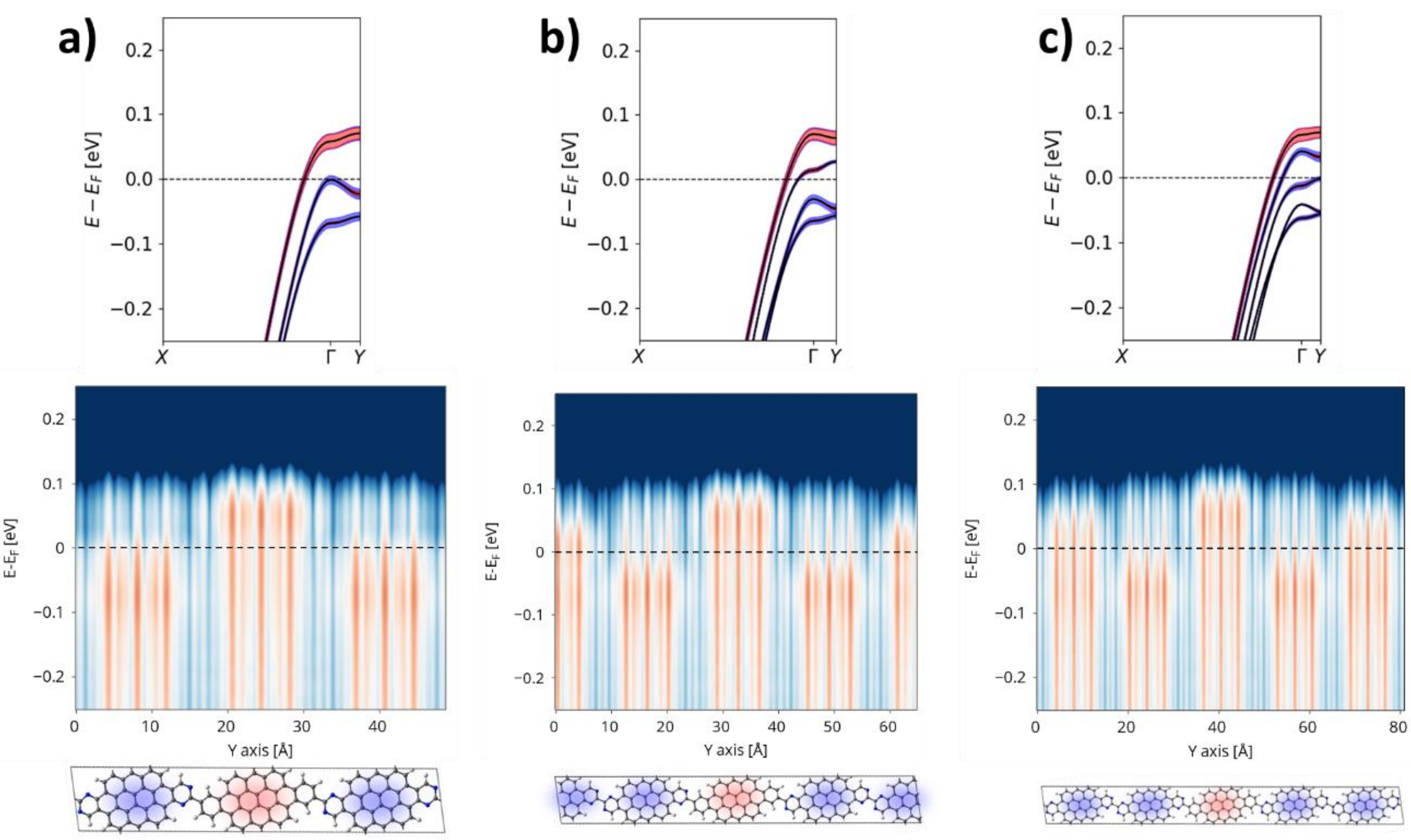

Figure S8. Electronic analysis of the -0.90 · $10^{13}$ e/cm$^{2}$ p-gated a) hNPG-1,2, b) hNPG-1,3 and c) hNPG-1,4 heterostructures. Top panels display the GNR-resolved fat band structure (cGNR: red; nGNR: blue) for each case. Middle panels display the y-resolved LDOS, including each corresponding heterostructure geometry at the bottom. The colors are in a log scale, with dark red (blue) indicating a high (low) LDOS. $E_F$ is indicated with a dashed black line.

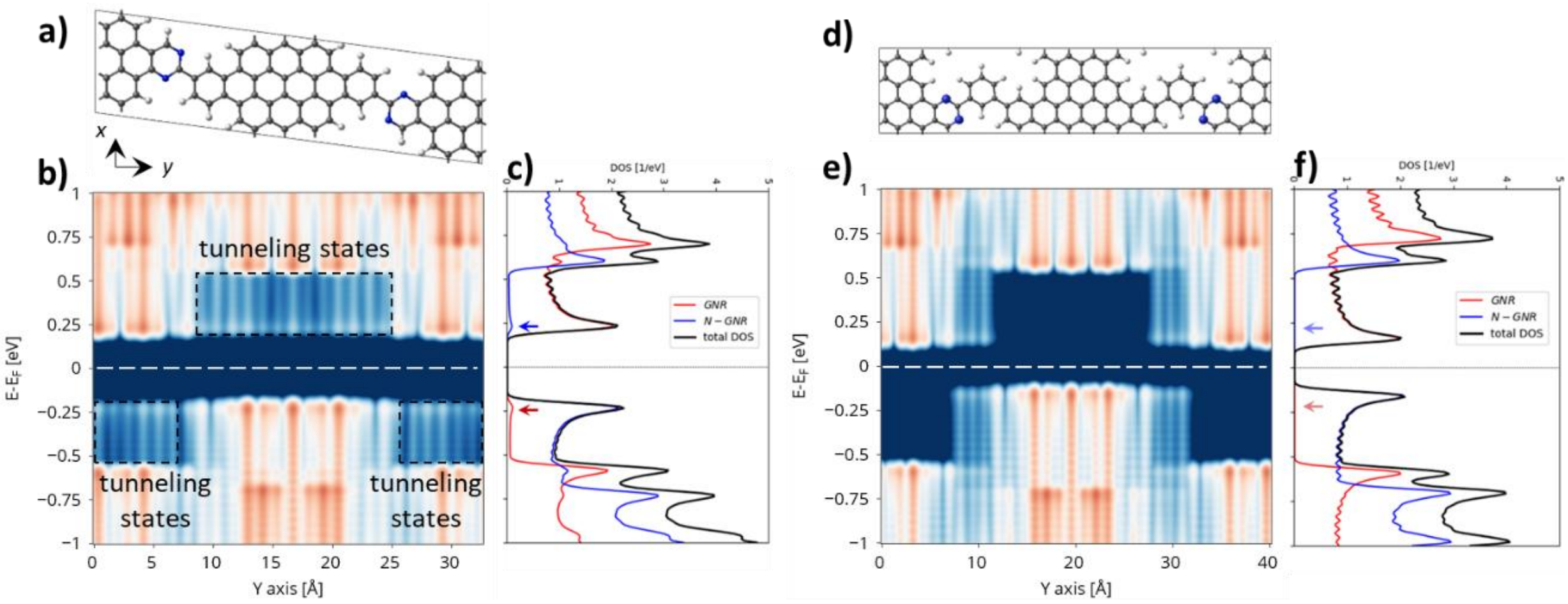

Figure S9. Comparison between the atomic structures, y-resolved LDOS and partial density of states (PDOS) for hNPG (a-c) and meta-hNPG (d-f). In panels c) and f) the cGNR and nGNR contributions are indicated with red and blue curves, respectively, with the total DOS shown in black. In the LDOS plots (b and e) the colors are in a log scale, with dark red (blue) indicating a high (low) LDOS. $E_F$ is indicated with a dashed white line.

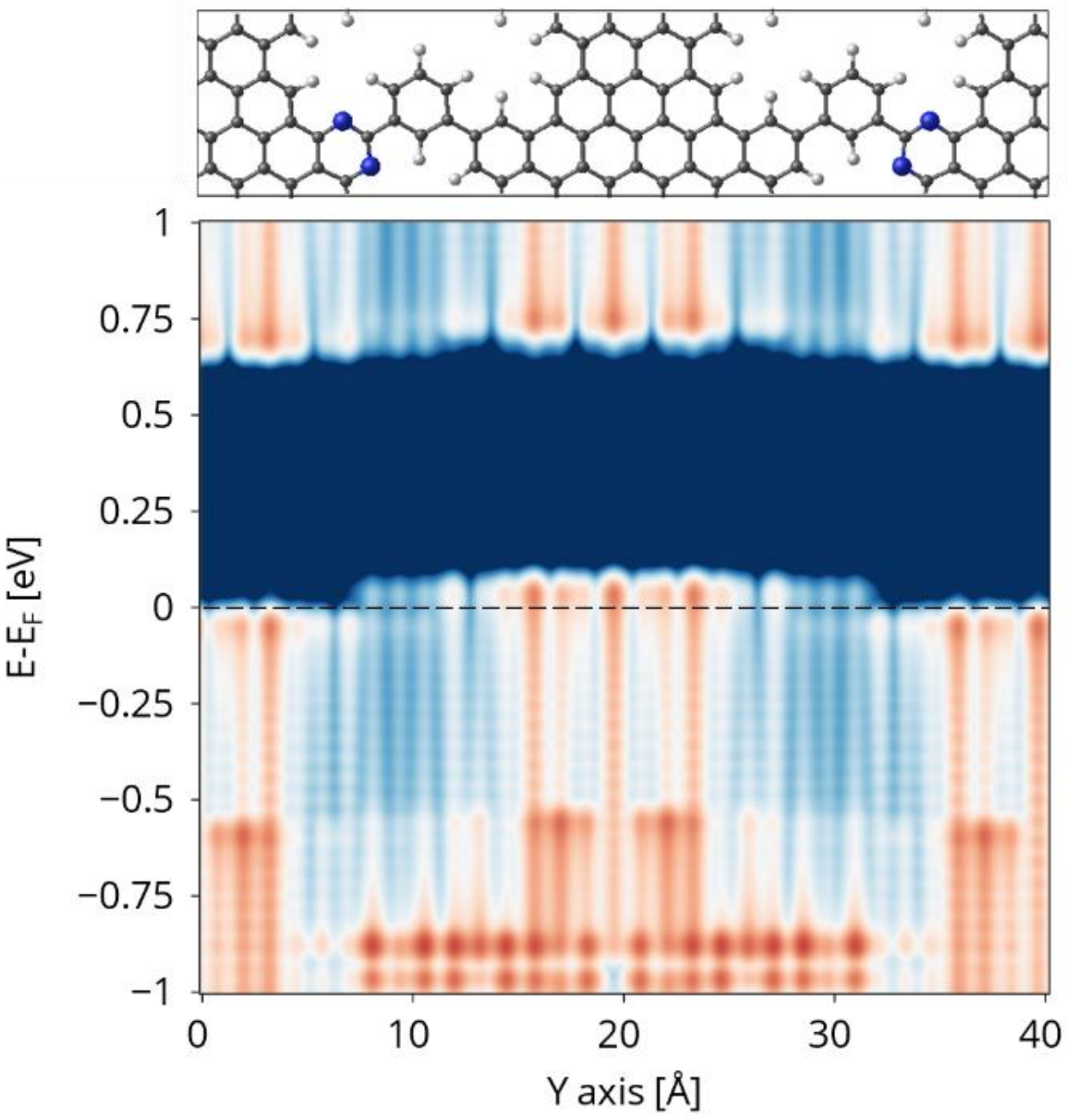

Figure S10. y-resolved LDOS for the -0.72 · $10^{13}$ e/$cm^2$ p-gated meta-hNPG (-0.25 e/cell). The colors are in log scale, with dark red (blue) indicating a high (low) LDOS. $E_F$ is indicated with a dashed black line. The meta-hNPG atomic structure is provided in the top panel for a one-to-one comparison with the LDOS plot.

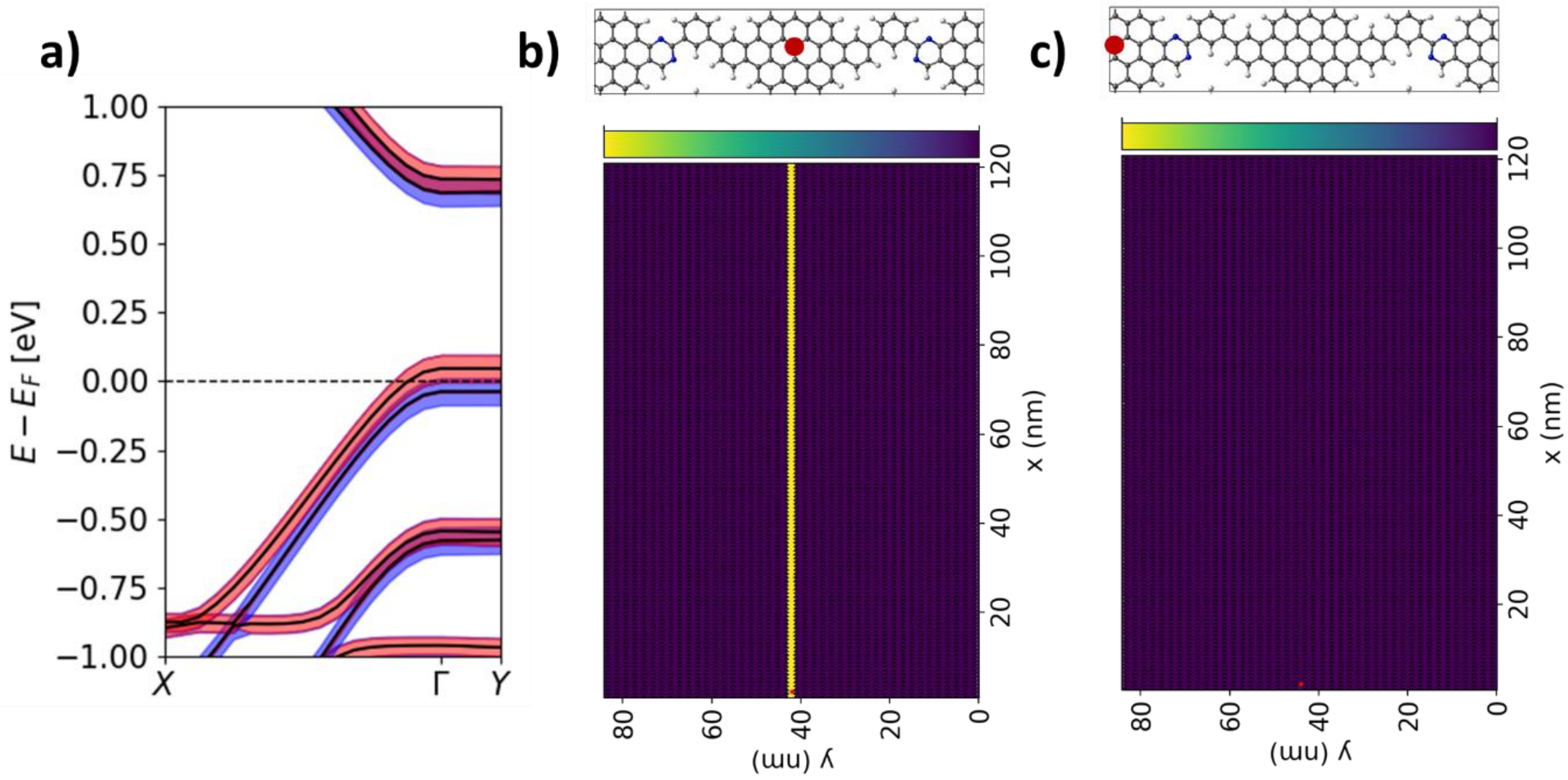

Figure S11. a) GNR-resolved fat band structure (cGNR: red; nGNR: blue) of meta-hNPG for -0.25 e/cell gating (-0.72 · $10^{13}$ e/$cm^2$) and corresponding bond transmission maps of a ~85 x 122 $nm^2$ sample upon locally contacting b) a cGNR or c) a nGNR at the bottom part of the sample (see small red dot. All color bars in the bond transmission maps range from 0 to 0.045. The Hamiltonian of these large-scale samples has been obtained by taking $p_z$ orbitals from the DFT Hamiltonian (see Methods for details). Top panels show the atomic structure of meta-hNPG, indicating with a red sphere the injection point in each case.

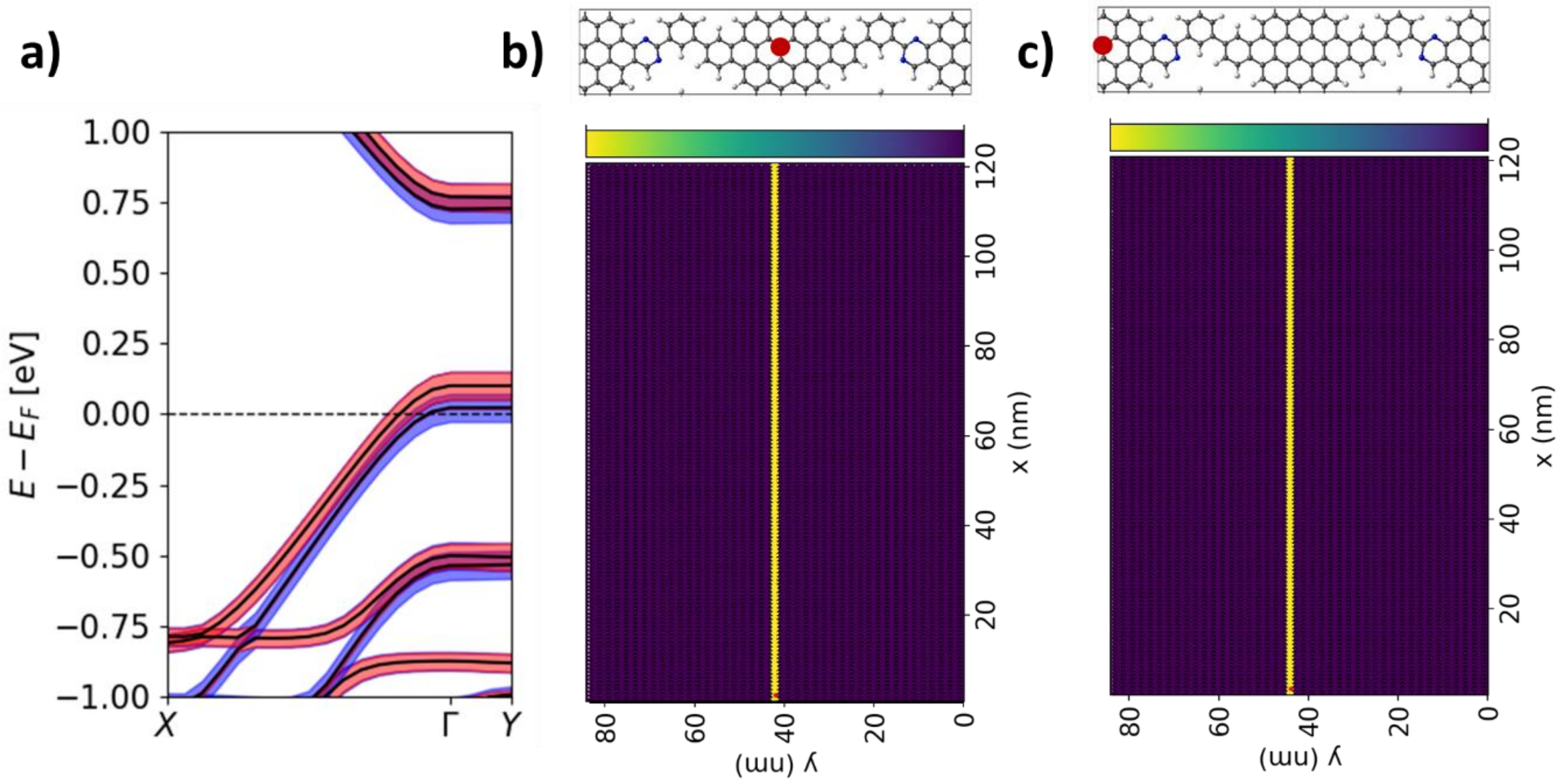

Figure S12. a) GNR-resolved fat band structure (cGNR: red; nGNR: blue) of meta-hNPG for -0.5 e/cell gating (-1.44 · $10^{13}$ e/cm$^2$) and corresponding bond transmission maps of a ~85 x 122 nm$^2$ sample upon locally contacting b) a cGNR or c) a nGNR at the bottom part of the sample (see small red dot). All color bars in the bond transmission maps range from 0 to 0.045. The Hamiltonian of these large-scale samples has been obtained by taking $p_z$ orbitals from the DFT Hamiltonian (see Methods for details). Top panels show the atomic structure of meta-hNPG, indicating with a red sphere the injection point in each case.

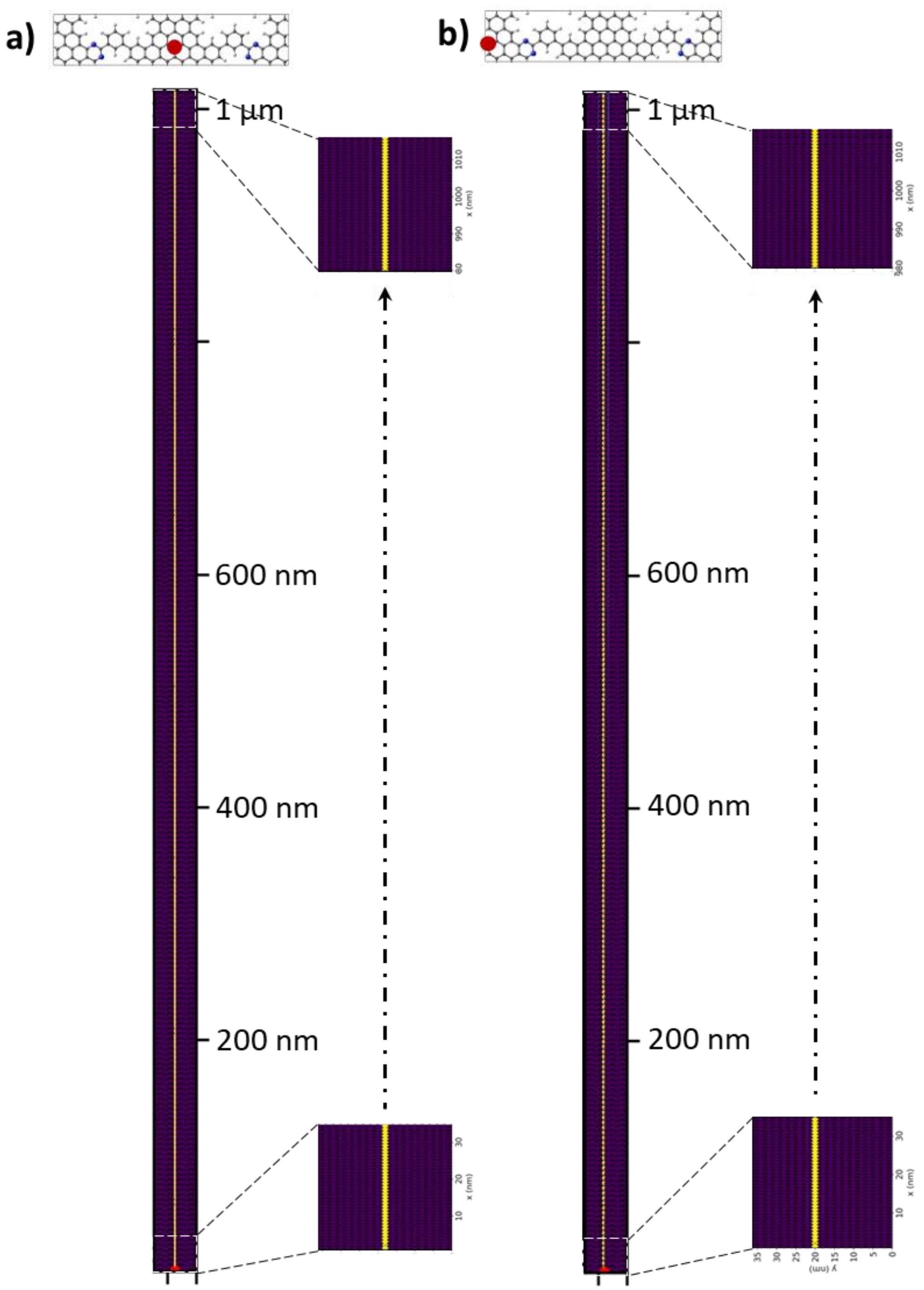

Figure S13. Bond transmission maps of a -0.5 e/cell gated ($-1.44 \cdot 10^{13}$ e/cm$^2$) ~36 x 1017 nm$^2$ meta-hNPG sample (968,760 atoms) upon locally contacting the bottom region of a) the central cGNR or b) the central nGNR (see red dots in the atomic structure in the top panel and in the device bottom region). A zoomed visual of the bond transmission maps is provided for the bottom and top regions, to highlight the level of collimation 1 µm away from the injection point. All color bars range from 0 to 0.045.

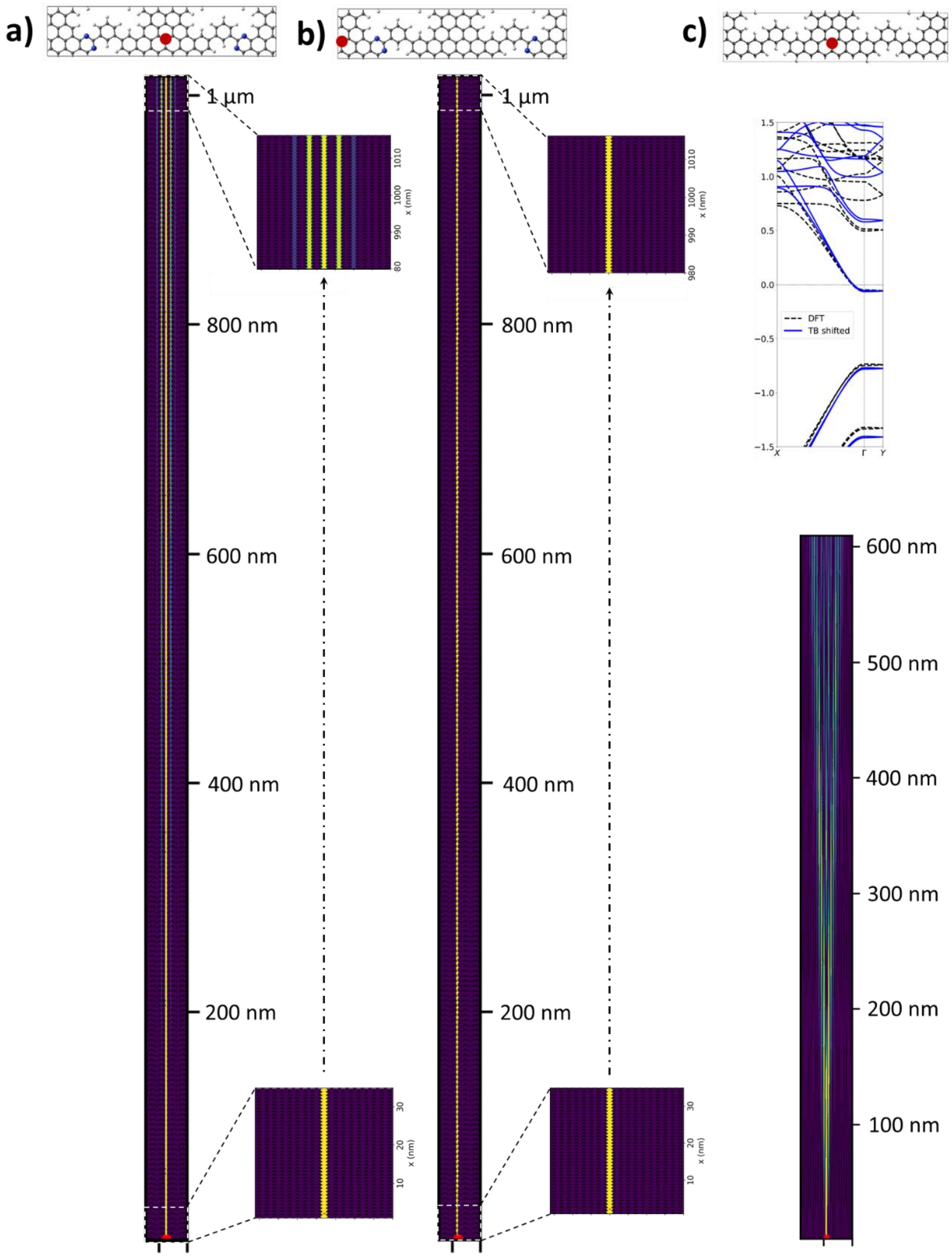

Figure S14 Bond transmission maps of a +0.5 e/cell gated ($+1.44 \cdot 10^{13}$ e/cm$^2$) ~36 x 1017 nm$^2$ meta-hNPG sample (968,760 atoms) upon locally contacting a) a cGNR or b) a nGNR at the bottom of the device (see small red dot). All color bars in the bond transmission maps range from 0 to 0.045. The injection point is shown as a red sphere in the atomic structures provided at the top. For both cases, a zoomed visual of the bond transmission map is provided for the bottom and top regions of the device, to highlight their different behavior far away from the source. c) Bond transmission maps of a +0.5 e/cell gated ($-1.42 \cdot 10^{13}$ e/cm$^2$) ~45 x 610 nm$^2$ meta-NPG sample (710,424 atoms) upon locally injecting at the bottom of the device (see small red dot). All color bars range from 0 to 0.045. The corresponding optimized atomic structure and band structure are provided in the top panels. Both DFT and DFT-parameterized TB bands are shown, with aligned conduction band edges (see Methods).

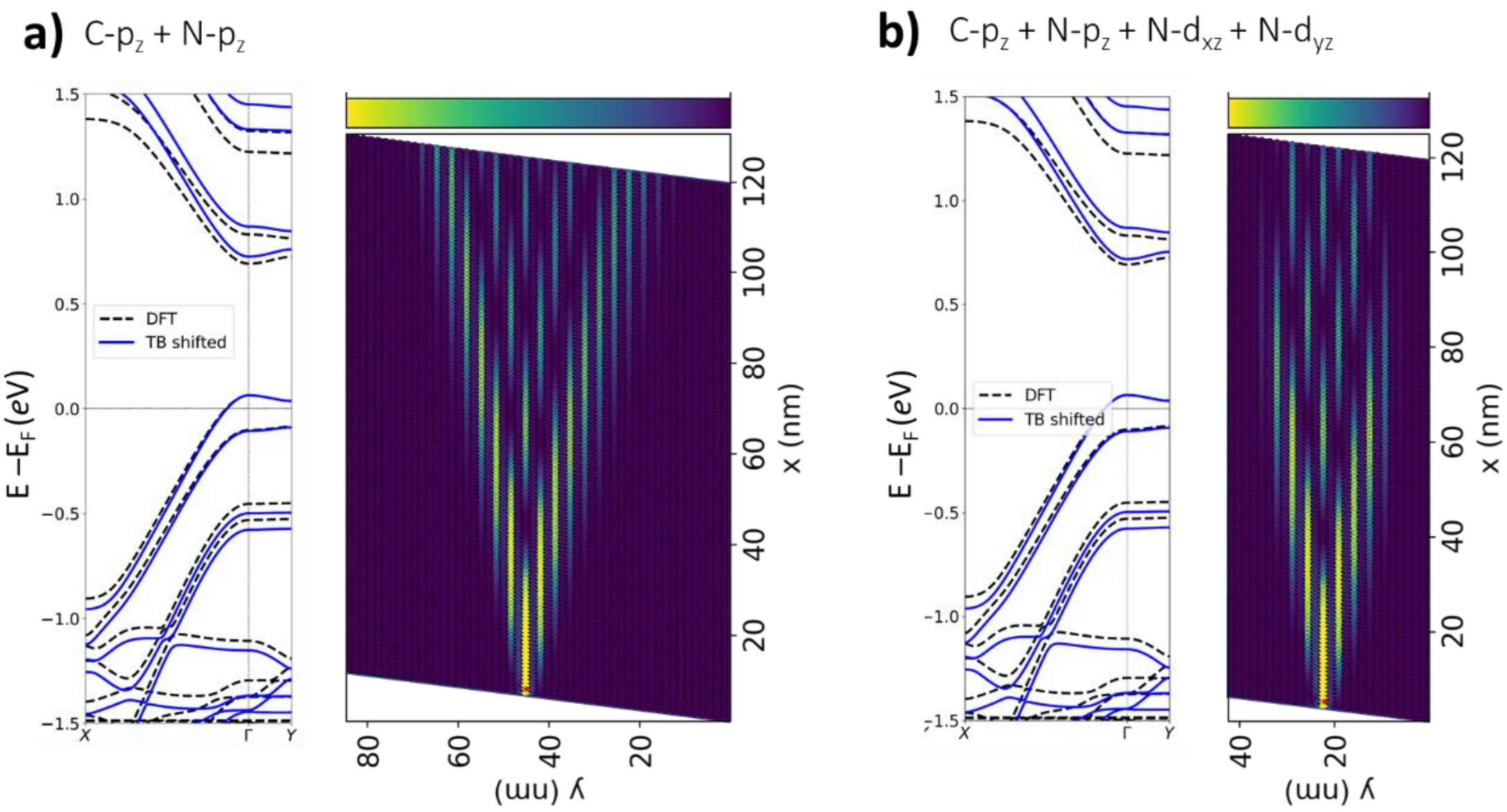

Figure S15 Comparison between band structures and bond transmission maps of a -0.25 e/cell p-gated hNPG system (-0.89 x $10^{13}$ e/cm$^3$) obtained by parameterizing the tight-binding Hamiltonian using DFT-converged Hamiltonian elements associated with (a) C-$p_z$ and N-$p_z$ or (b) C-$p_z$, N-$p_z$, N-$d_{xz}$ and N$d_{yz}$ orbitals. All color bars in the bond transmission maps range from 0 to 0.045.

## Supplementary Note S1

To validate the model based on coupled discrete differential equations (DDE), we compare the field intensity obtained from a parameterized model with bond-resolved transmissions obtained with the TB-GF methodology. The parametrization for the propagation constant ($\alpha$) was set to:

$$\alpha = \frac{U}{\overline{dE/dk}} = \frac{|\overline{\epsilon_C} - \overline{\epsilon_N}|}{\overline{dE/dk}}.$$

This provides a good estimate of the model parameters of the model and establishes a connection between the propagation along the waveguides in the DDE model with the difference in the average $p_z$ on-site potentials of cGNRs/nGNRs ($|\overline{\epsilon_C} - \overline{\epsilon_N}|$), normalized by the average band slope of the conduction band (group velocity of the conducting electrons, $\overline{dE/dk}$). For an n-doped hNPG device ($0.36 \cdot 10^{13}$ e/cm$^2$), we get $U = 0.248\ eV$ and an average band slope of $3.81\ eV/\text{Å}$. With this, $\alpha = 0.065\ \text{Å}^{-1}$was obtained. We set $\kappa = 0.0137\ \text{Å}^{-1}$ from[1]

$$\kappa = \frac{|k_1 - k_0|}{4},$$

with $k_0$ and $k_1$ being the momenta associated with the first two conduction (or valence) linear bands dispersing along the Γ-X high-symmetry path of the Brillouin zone, evaluated at the injection energy $E$ of $E - E_F = 0.1\ eV$. The TB-GF device used for comparison consists of an n-gated hNPG device ($0.36 \cdot 10^{13}$ e/cm$^2$) of 68.9 nm × 103 nm with 252,000 atoms, with the injection site located on a nGNR. As shown in Figure S16, we find good qualitative agreement between the field intensity propagation of the DDE model and the large-scale TB-GF calculations. The latter exhibits faster oscillations, attributed to intra-unit-cell hopping processes, which are not captured by the coarse-grained nature of our effective model (Figure S17b). Additionally, the amplitude of the oscillations is larger in the TB-GF results (Figure S17b), likely due to local variations in the on-site $p_z$ orbital energies along the GNRs, as well as different atomic sublattices in the hNPG unit-cell (Figure S17a).

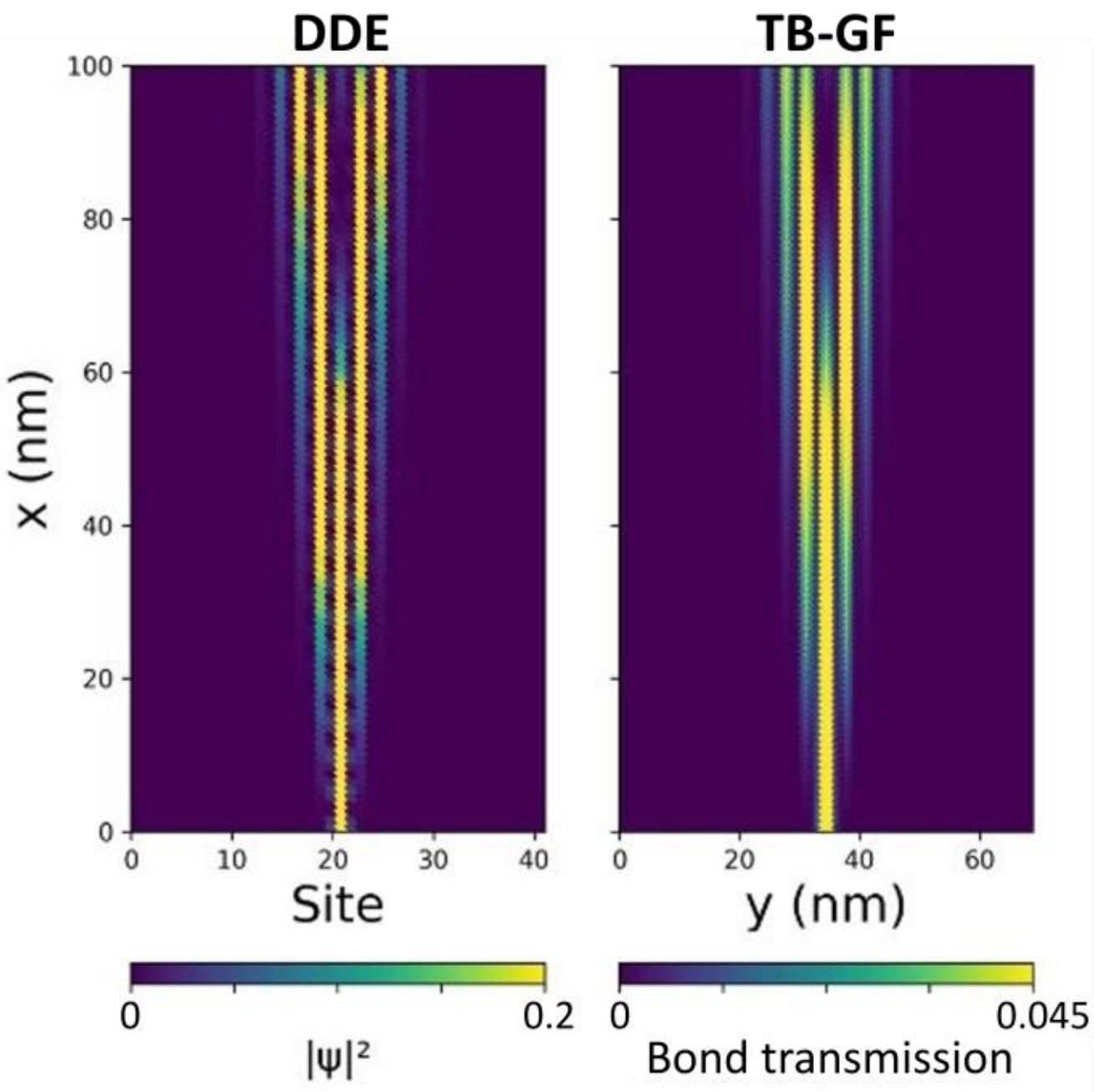

Figure S16. Comparison between the spatial intensity of waveguide modes from the DDE model and the bond transmission of large-scale TB-GF transport calculations for an n-gated hNPG device ($0.36 \cdot 10^{13}$ e/cm$^2$) contacted on an nGNR. Here $\alpha$=0.065 Å$^{-1}$ and $\kappa$=0.014 Å$^{-1}$ (see Methods).

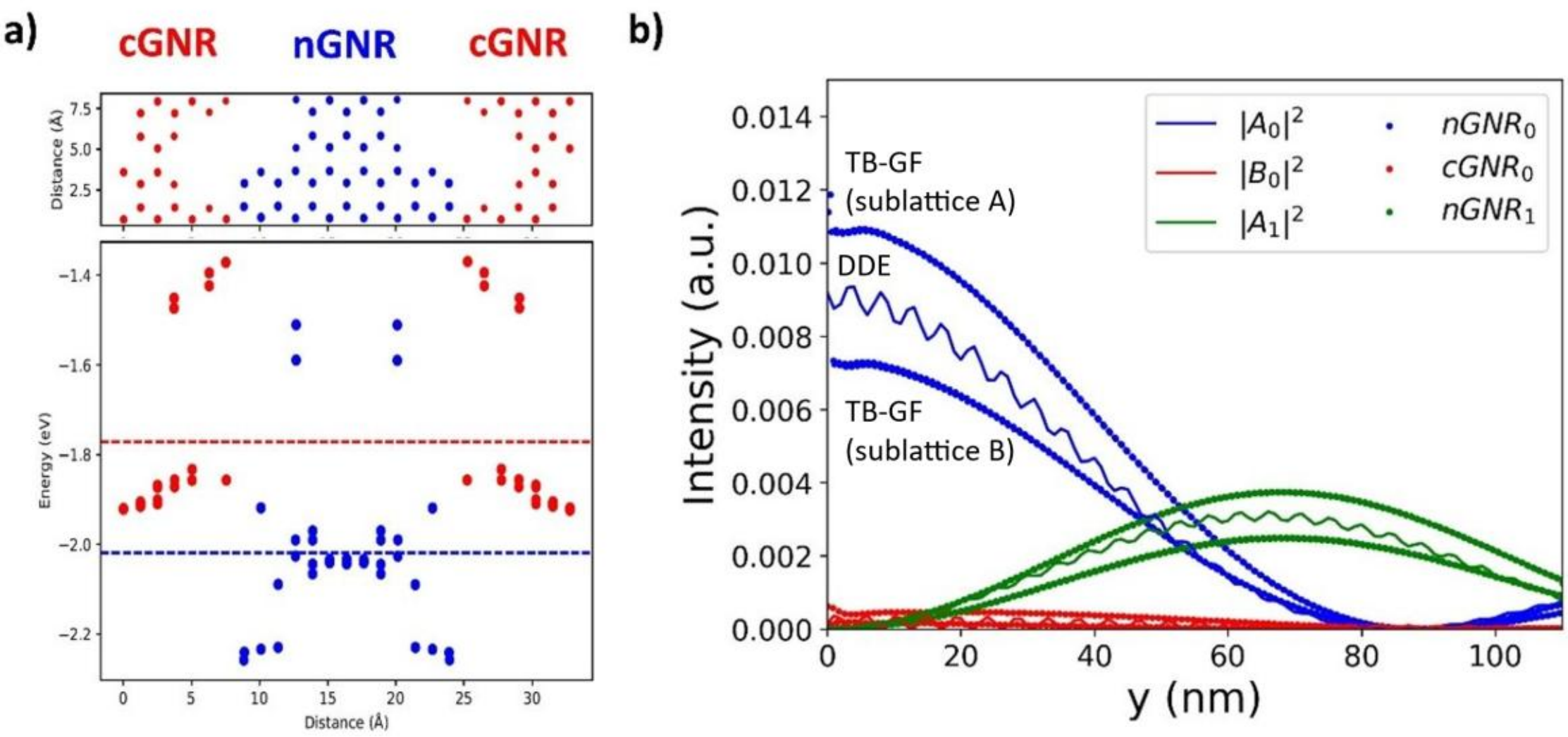

Figure S17. Parametrization of the DDE model for an n doped hNPG ($0.36 \cdot 10^{13}$ e/cm$^2$). (a) The top panel shows the hNPG structure from the parameter-free TB. The coloring of the atoms corresponds to the type of subsystem they belong to, with cGNR/nGNR being red/blue. The carbon $p_z$ onsite energies are shown in the bottom panel. The dashed lines represent the average for each GNR used to determine $\alpha$. (b) Comparison between the waveguide mode intensity from the DDE model (line) and the atom transmission from TB-GF calculations at the center of the GNRs (dotted) for the first three GNRs. In the TB-GF case, two dotted lines for each GNR illustrate the contributions from the two graphene sublattices. The waveguide mode intensity from the DDE model was scaled for comparison. The injection point in the TB-GF calculations was on an nGNR, which is conducting for the gating considered here.

Finally, we would like to note that the DFT calculations, by which the DDE model has been based upon, do not include electron-electron correlation effects, which have been shown to potentially reduce the band gap of carbon nanomaterials.[2] However, as demonstrated in that theoretical study, as long as the materials band gap is much larger than kBT (as in our case), electron-electron scattering should remain low, and so our parametrized $\alpha$ and $\kappa$ values should not qualitatively change.

Robustness against random variations in the effective parameter $\alpha$, up to $0.05\ \alpha$, at individual ribbons shows that the interference pattern persists with statistic disorder for instance from thermal fluctuations.

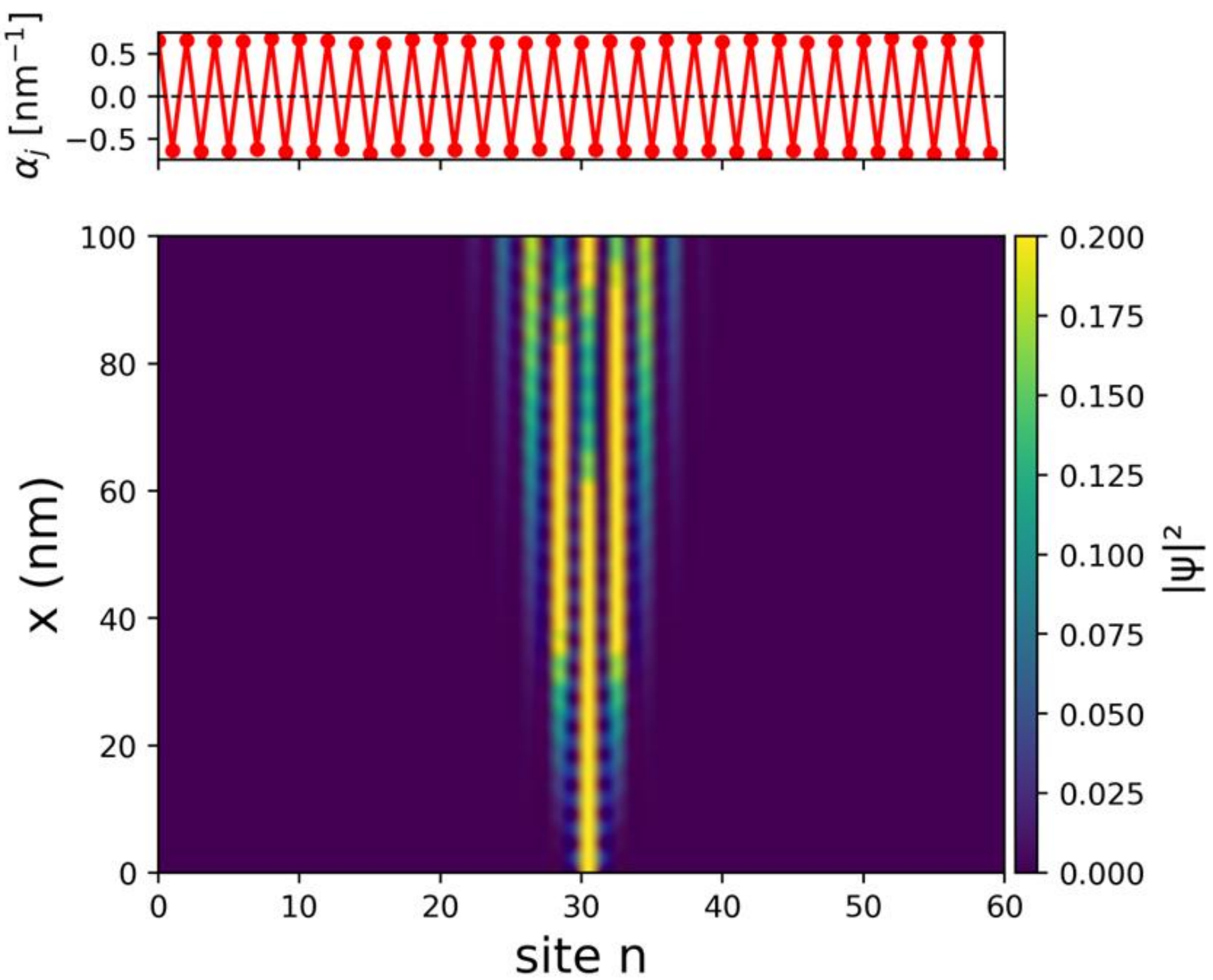

Figure S18. Spatial intensity of waveguide modes from the DDE model under random variations. Here $\alpha_j=(-1)^j\ 0.065+\delta_j$ Å$^{-1}$ and $\kappa=0.014$ Å$^{-1}$ (see Methods).